\documentclass[longauth]{aa}

\usepackage{graphicx}
\usepackage{natbib}
\usepackage{scalerel}

\usepackage{xcolor}

\usepackage{txfonts}
\usepackage[pdfencoding=auto,psdextra]{hyperref}
\hypersetup{
    colorlinks=true,
    linkcolor=blue,
    filecolor=magenta,      
    urlcolor=blue,
    citecolor=blue
}
\makeatletter
\renewcommand*\aa@pageof{, page \thepage{} of \pageref*{LastPage}}
\makeatother

\usepackage[utf8]{inputenc}

\usepackage[switch, modulo]{lineno}
\usepackage{orcidlink} %this has to come after hyperref
\newcommand{\orcid}[1]{\orcidlink{#1}}
\usepackage{euclid}

\begin{document}
%
% Put the title of your paper here:
%
 %  \title{\Euclid: Early Release Observations --  a glance at free-floating planets in the $\sigma$\,Orionis cluster
   %I. 
  % A deep glimpse of Orion and Taurus. 
 % Barnard 33, NGC\,2023, IC\,434 and the $\sigma$\,Orionis open cluster
 \title{\Euclid: Early Release Observations -- A glance at free-floating new-born planets in the $\sigma$\,Orionis cluster\thanks{This paper is published on behalf of the Euclid Consortium.}}    
 %  \thanks{This paper is published on
  %   behalf of the Euclid Collaboration}}

%%%% please do not edit the author list -- contact ECEB Bureau for changes
%\newcommand{\orcid}[1]{} %% if already defined in aa.cls: comment, or use renewcommand			   
\author{E.~L.~Mart\'in\orcid{0000-0002-1208-4833}\thanks{\email{ege@iac.es}}\inst{\ref{aff1},\ref{aff2}}
\and M.~{\v Z}erjal\orcid{0000-0001-6023-4974}\inst{\ref{aff1},\ref{aff2}}
\and H.~Bouy\inst{\ref{aff3},\ref{aff4}}
\and D.~Martin-Gonzalez\orcid{0009-0001-6964-1880}\inst{\ref{aff5}}
\and S.~Mu{\~ n}oz~Torres\orcid{0000-0003-4269-4779}\inst{\ref{aff1},\ref{aff2}}
\and D.~Barrado\orcid{0000-0002-5971-9242}\inst{\ref{aff6}}
\and J.~Olivares\orcid{0000-0003-0316-2956}\inst{\ref{aff7}}
\and A.~P\'erez-Garrido\orcid{0000-0002-5139-1975}\inst{\ref{aff8}}
\and P.~Mas-Buitrago\orcid{0000-0001-8055-7949}\inst{\ref{aff6}}
\and P.~Cruz\orcid{0000-0003-1793-200X}\inst{\ref{aff6}}
\and E.~Solano\inst{\ref{aff6}}
\and M.~R.~Zapatero~Osorio\orcid{0000-0001-5664-2852}\inst{\ref{aff6}}
\and N.~Lodieu\orcid{0000-0002-3612-8968}\inst{\ref{aff1},\ref{aff2}}
\and V.~J.~S.~B\'ejar\inst{\ref{aff1},\ref{aff2}}
\and J.-Y.~Zhang\orcid{0000-0001-5392-2701}\inst{\ref{aff1},\ref{aff2}}
\and C.~del~Burgo\orcid{0000-0002-8949-5200}\inst{\ref{aff1},\ref{aff2}}
\and N.~Hu\'elamo \inst{\ref{aff6}}
\and R.~Laureijs\inst{\ref{aff9}}
\and A.~Mora\orcid{0000-0002-1922-8529}\inst{\ref{aff10}}
\and T.~Saifollahi\orcid{0000-0002-9554-7660}\inst{\ref{aff11},\ref{aff12}}
\and J.-C.~Cuillandre\orcid{0000-0002-3263-8645}\inst{\ref{aff13}}
\and M.~Schirmer\orcid{0000-0003-2568-9994}\inst{\ref{aff14}}
\and R.~Tata\inst{\ref{aff15}}
\and S.~Points\orcid{0000-0002-4596-1337}\inst{\ref{aff16}}
\and N.~Phan-Bao\orcid{0000-0002-4481-0094}\inst{\ref{aff17},\ref{aff18}}
\and B.~Goldman\orcid{0000-0002-2729-7276}\inst{\ref{aff19},\ref{aff11}}
\and S.~L.~Casewell\orcid{0000-0003-2478-0120}\inst{\ref{aff20}}
\and C.~Reyl\'e\orcid{0000-0003-2258-2403}\inst{\ref{aff21}}
\and R.~L.~Smart\orcid{0000-0002-4424-4766}\inst{\ref{aff22},\ref{aff23}}
\and N.~Aghanim\orcid{0000-0002-6688-8992}\inst{\ref{aff24}}
\and B.~Altieri\orcid{0000-0003-3936-0284}\inst{\ref{aff25}}
\and S.~Andreon\orcid{0000-0002-2041-8784}\inst{\ref{aff26}}
\and N.~Auricchio\orcid{0000-0003-4444-8651}\inst{\ref{aff27}}
\and M.~Baldi\orcid{0000-0003-4145-1943}\inst{\ref{aff28},\ref{aff27},\ref{aff29}}
\and A.~Balestra\orcid{0000-0002-6967-261X}\inst{\ref{aff30}}
\and S.~Bardelli\orcid{0000-0002-8900-0298}\inst{\ref{aff27}}
\and A.~Basset\inst{\ref{aff31}}
\and R.~Bender\orcid{0000-0001-7179-0626}\inst{\ref{aff32},\ref{aff33}}
\and D.~Bonino\orcid{0000-0002-3336-9977}\inst{\ref{aff22}}
\and E.~Branchini\orcid{0000-0002-0808-6908}\inst{\ref{aff34},\ref{aff35},\ref{aff26}}
\and M.~Brescia\orcid{0000-0001-9506-5680}\inst{\ref{aff36},\ref{aff37},\ref{aff38}}
\and J.~Brinchmann\orcid{0000-0003-4359-8797}\inst{\ref{aff39}}
\and S.~Camera\orcid{0000-0003-3399-3574}\inst{\ref{aff40},\ref{aff41},\ref{aff22}}
\and V.~Capobianco\orcid{0000-0002-3309-7692}\inst{\ref{aff22}}
\and C.~Carbone\orcid{0000-0003-0125-3563}\inst{\ref{aff42}}
\and J.~Carretero\orcid{0000-0002-3130-0204}\inst{\ref{aff43},\ref{aff44}}
\and S.~Casas\orcid{0000-0002-4751-5138}\inst{\ref{aff45}}
\and M.~Castellano\orcid{0000-0001-9875-8263}\inst{\ref{aff46}}
\and S.~Cavuoti\orcid{0000-0002-3787-4196}\inst{\ref{aff37},\ref{aff38}}
\and A.~Cimatti\inst{\ref{aff47}}
\and G.~Congedo\orcid{0000-0003-2508-0046}\inst{\ref{aff48}}
\and C.~J.~Conselice\orcid{0000-0003-1949-7638}\inst{\ref{aff49}}
\and L.~Conversi\orcid{0000-0002-6710-8476}\inst{\ref{aff50},\ref{aff25}}
\and Y.~Copin\orcid{0000-0002-5317-7518}\inst{\ref{aff51}}
\and L.~Corcione\orcid{0000-0002-6497-5881}\inst{\ref{aff22}}
\and F.~Courbin\orcid{0000-0003-0758-6510}\inst{\ref{aff52}}
\and H.~M.~Courtois\orcid{0000-0003-0509-1776}\inst{\ref{aff53}}
\and M.~Cropper\orcid{0000-0003-4571-9468}\inst{\ref{aff54}}
\and A.~Da~Silva\orcid{0000-0002-6385-1609}\inst{\ref{aff55},\ref{aff56}}
\and H.~Degaudenzi\orcid{0000-0002-5887-6799}\inst{\ref{aff57}}
\and A.~M.~Di~Giorgio\orcid{0000-0002-4767-2360}\inst{\ref{aff58}}
\and J.~Dinis\orcid{0000-0001-5075-1601}\inst{\ref{aff55},\ref{aff56}}
\and F.~Dubath\orcid{0000-0002-6533-2810}\inst{\ref{aff57}}
\and X.~Dupac\inst{\ref{aff25}}
\and S.~Dusini\orcid{0000-0002-1128-0664}\inst{\ref{aff59}}
\and A.~Ealet\orcid{0000-0003-3070-014X}\inst{\ref{aff51}}
\and M.~Farina\orcid{0000-0002-3089-7846}\inst{\ref{aff58}}
\and S.~Farrens\orcid{0000-0002-9594-9387}\inst{\ref{aff13}}
\and S.~Ferriol\inst{\ref{aff51}}
\and P.~Fosalba\orcid{0000-0002-1510-5214}\inst{\ref{aff60},\ref{aff61}}
\and M.~Frailis\orcid{0000-0002-7400-2135}\inst{\ref{aff62}}
\and E.~Franceschi\orcid{0000-0002-0585-6591}\inst{\ref{aff27}}
\and M.~Fumana\orcid{0000-0001-6787-5950}\inst{\ref{aff42}}
\and S.~Galeotta\orcid{0000-0002-3748-5115}\inst{\ref{aff62}}
\and B.~Garilli\orcid{0000-0001-7455-8750}\inst{\ref{aff42}}
\and W.~Gillard\orcid{0000-0003-4744-9748}\inst{\ref{aff63}}
\and B.~Gillis\orcid{0000-0002-4478-1270}\inst{\ref{aff48}}
\and C.~Giocoli\orcid{0000-0002-9590-7961}\inst{\ref{aff27},\ref{aff64}}
\and P.~G\'omez-Alvarez\orcid{0000-0002-8594-5358}\inst{\ref{aff65},\ref{aff25}}
\and A.~Grazian\orcid{0000-0002-5688-0663}\inst{\ref{aff30}}
\and F.~Grupp\inst{\ref{aff32},\ref{aff33}}
\and L.~Guzzo\orcid{0000-0001-8264-5192}\inst{\ref{aff66},\ref{aff26}}
\and S.~V.~H.~Haugan\orcid{0000-0001-9648-7260}\inst{\ref{aff67}}
\and J.~Hoar\inst{\ref{aff25}}
\and H.~Hoekstra\orcid{0000-0002-0641-3231}\inst{\ref{aff68}}
\and W.~Holmes\inst{\ref{aff69}}
\and I.~Hook\orcid{0000-0002-2960-978X}\inst{\ref{aff70}}
\and F.~Hormuth\inst{\ref{aff71}}
\and A.~Hornstrup\orcid{0000-0002-3363-0936}\inst{\ref{aff72},\ref{aff73}}
\and D.~Hu\inst{\ref{aff54}}
\and P.~Hudelot\inst{\ref{aff74}}
\and K.~Jahnke\orcid{0000-0003-3804-2137}\inst{\ref{aff14}}
\and M.~Jhabvala\inst{\ref{aff75}}
\and E.~Keih\"anen\orcid{0000-0003-1804-7715}\inst{\ref{aff76}}
\and S.~Kermiche\orcid{0000-0002-0302-5735}\inst{\ref{aff63}}
\and A.~Kiessling\orcid{0000-0002-2590-1273}\inst{\ref{aff69}}
\and M.~Kilbinger\orcid{0000-0001-9513-7138}\inst{\ref{aff77}}
\and T.~Kitching\orcid{0000-0002-4061-4598}\inst{\ref{aff54}}
\and R.~Kohley\inst{\ref{aff25}}
\and B.~Kubik\orcid{0009-0006-5823-4880}\inst{\ref{aff51}}
\and M.~K\"ummel\orcid{0000-0003-2791-2117}\inst{\ref{aff33}}
\and M.~Kunz\orcid{0000-0002-3052-7394}\inst{\ref{aff78}}
\and H.~Kurki-Suonio\orcid{0000-0002-4618-3063}\inst{\ref{aff79},\ref{aff80}}
\and D.~Le~Mignant\orcid{0000-0002-5339-5515}\inst{\ref{aff81}}
\and S.~Ligori\orcid{0000-0003-4172-4606}\inst{\ref{aff22}}
\and P.~B.~Lilje\orcid{0000-0003-4324-7794}\inst{\ref{aff67}}
\and V.~Lindholm\orcid{0000-0003-2317-5471}\inst{\ref{aff79},\ref{aff80}}
\and I.~Lloro\inst{\ref{aff82}}
\and D.~Maino\inst{\ref{aff66},\ref{aff42},\ref{aff83}}
\and E.~Maiorano\orcid{0000-0003-2593-4355}\inst{\ref{aff27}}
\and O.~Mansutti\orcid{0000-0001-5758-4658}\inst{\ref{aff62}}
\and O.~Marggraf\orcid{0000-0001-7242-3852}\inst{\ref{aff84}}
\and N.~Martinet\orcid{0000-0003-2786-7790}\inst{\ref{aff81}}
\and F.~Marulli\orcid{0000-0002-8850-0303}\inst{\ref{aff85},\ref{aff27},\ref{aff29}}
\and R.~Massey\orcid{0000-0002-6085-3780}\inst{\ref{aff86}}
\and E.~Medinaceli\orcid{0000-0002-4040-7783}\inst{\ref{aff27}}
\and S.~Mei\orcid{0000-0002-2849-559X}\inst{\ref{aff87}}
\and M.~Melchior\inst{\ref{aff88}}
\and Y.~Mellier\inst{\ref{aff89},\ref{aff74}}
\and M.~Meneghetti\orcid{0000-0003-1225-7084}\inst{\ref{aff27},\ref{aff29}}
\and G.~Meylan\inst{\ref{aff52}}
\and J.~J.~Mohr\orcid{0000-0002-6875-2087}\inst{\ref{aff33},\ref{aff32}}
\and M.~Moresco\orcid{0000-0002-7616-7136}\inst{\ref{aff85},\ref{aff27}}
\and L.~Moscardini\orcid{0000-0002-3473-6716}\inst{\ref{aff85},\ref{aff27},\ref{aff29}}
\and S.-M.~Niemi\inst{\ref{aff9}}
\and C.~Padilla\orcid{0000-0001-7951-0166}\inst{\ref{aff90}}
\and S.~Paltani\orcid{0000-0002-8108-9179}\inst{\ref{aff57}}
\and F.~Pasian\orcid{0000-0002-4869-3227}\inst{\ref{aff62}}
\and K.~Pedersen\inst{\ref{aff91}}
\and W.~J.~Percival\orcid{0000-0002-0644-5727}\inst{\ref{aff92},\ref{aff93},\ref{aff94}}
\and V.~Pettorino\inst{\ref{aff9}}
\and S.~Pires\orcid{0000-0002-0249-2104}\inst{\ref{aff13}}
\and G.~Polenta\orcid{0000-0003-4067-9196}\inst{\ref{aff95}}
\and M.~Poncet\inst{\ref{aff31}}
\and L.~A.~Popa\inst{\ref{aff96}}
\and L.~Pozzetti\orcid{0000-0001-7085-0412}\inst{\ref{aff27}}
\and G.~D.~Racca\inst{\ref{aff9}}
\and F.~Raison\orcid{0000-0002-7819-6918}\inst{\ref{aff32}}
\and R.~Rebolo\inst{\ref{aff1},\ref{aff2}}
\and A.~Renzi\orcid{0000-0001-9856-1970}\inst{\ref{aff97},\ref{aff59}}
\and J.~Rhodes\orcid{0000-0002-4485-8549}\inst{\ref{aff69}}
\and G.~Riccio\inst{\ref{aff37}}
\and Hans-Walter~Rix\orcid{0000-0003-4996-9069}\inst{\ref{aff14}}
\and E.~Romelli\orcid{0000-0003-3069-9222}\inst{\ref{aff62}}
\and M.~Roncarelli\orcid{0000-0001-9587-7822}\inst{\ref{aff27}}
\and E.~Rossetti\orcid{0000-0003-0238-4047}\inst{\ref{aff28}}
\and R.~Saglia\orcid{0000-0003-0378-7032}\inst{\ref{aff33},\ref{aff32}}
\and D.~Sapone\orcid{0000-0001-7089-4503}\inst{\ref{aff98}}
\and B.~Sartoris\orcid{0000-0003-1337-5269}\inst{\ref{aff33},\ref{aff62}}
\and M.~Sauvage\orcid{0000-0002-0809-2574}\inst{\ref{aff13}}
\and R.~Scaramella\orcid{0000-0003-2229-193X}\inst{\ref{aff46},\ref{aff99}}
\and P.~Schneider\orcid{0000-0001-8561-2679}\inst{\ref{aff84}}
\and A.~Secroun\orcid{0000-0003-0505-3710}\inst{\ref{aff63}}
\and G.~Seidel\orcid{0000-0003-2907-353X}\inst{\ref{aff14}}
\and M.~Seiffert\orcid{0000-0002-7536-9393}\inst{\ref{aff69}}
\and S.~Serrano\orcid{0000-0002-0211-2861}\inst{\ref{aff60},\ref{aff100},\ref{aff101}}
\and C.~Sirignano\orcid{0000-0002-0995-7146}\inst{\ref{aff97},\ref{aff59}}
\and G.~Sirri\orcid{0000-0003-2626-2853}\inst{\ref{aff29}}
\and L.~Stanco\orcid{0000-0002-9706-5104}\inst{\ref{aff59}}
\and P.~Tallada-Cresp\'{i}\orcid{0000-0002-1336-8328}\inst{\ref{aff43},\ref{aff44}}
\and A.~N.~Taylor\inst{\ref{aff48}}
\and H.~I.~Teplitz\orcid{0000-0002-7064-5424}\inst{\ref{aff102}}
\and I.~Tereno\inst{\ref{aff55},\ref{aff103}}
\and R.~Toledo-Moreo\orcid{0000-0002-2997-4859}\inst{\ref{aff104}}
\and A.~Tsyganov\inst{\ref{aff105}}
\and I.~Tutusaus\orcid{0000-0002-3199-0399}\inst{\ref{aff106}}
\and L.~Valenziano\orcid{0000-0002-1170-0104}\inst{\ref{aff27},\ref{aff107}}
\and T.~Vassallo\orcid{0000-0001-6512-6358}\inst{\ref{aff33},\ref{aff62}}
\and G.~Verdoes~Kleijn\orcid{0000-0001-5803-2580}\inst{\ref{aff12}}
\and Y.~Wang\orcid{0000-0002-4749-2984}\inst{\ref{aff102}}
\and J.~Weller\orcid{0000-0002-8282-2010}\inst{\ref{aff33},\ref{aff32}}
\and O.~R.~Williams\orcid{0000-0003-0274-1526}\inst{\ref{aff105}}
\and E.~Zucca\orcid{0000-0002-5845-8132}\inst{\ref{aff27}}
\and C.~Baccigalupi\orcid{0000-0002-8211-1630}\inst{\ref{aff108},\ref{aff62},\ref{aff109},\ref{aff110}}
\and G.~Willis\inst{\ref{aff54}}
\and P.~Simon\inst{\ref{aff84}}
\and J.~Mart\'{i}n-Fleitas\orcid{0000-0002-8594-569X}\inst{\ref{aff10}}
\and D.~Scott\orcid{0000-0002-6878-9840}\inst{\ref{aff111}}}
										   
%%%% please do not edit the affiliation list -- contact ECEB Bureau for changes
\institute{Instituto de Astrof\'isica de Canarias, Calle V\'ia L\'actea s/n, 38204, San Crist\'obal de La Laguna, Tenerife, Spain\label{aff1}
\and
Departamento de Astrof\'isica, Universidad de La Laguna, 38206, La Laguna, Tenerife, Spain\label{aff2}
\and
Laboratoire d'Astrophysique de Bordeaux, CNRS and Universit\'e de Bordeaux, All\'ee Geoffroy St. Hilaire, 33165 Pessac, France\label{aff3}
\and
Institut universitaire de France (IUF), 1 rue Descartes, 75231 PARIS CEDEX 05, France\label{aff4}
\and
Departamento de F{\'\i}sica Fundamental. Universidad de Salamanca. Plaza de la Merced s/n. 37008 Salamanca, Spain\label{aff5}
\and
Centro de Astrobiolog\'ia (CAB), CSIC-INTA, ESAC Campus, Camino Bajo del Castillo s/n, 28692 Villanueva de la Ca\~nada, Madrid, Spain\label{aff6}
\and
Departamento de Inteligencia Artificial, Universidad Nacional de Educaci\'on a Distancia (UNED), c/Juan del Rosal 16, E-28040, Madrid, Spain\label{aff7}
\and
Departamento F\'isica Aplicada, Universidad Polit\'ecnica de Cartagena, Campus Muralla del Mar, 30202 Cartagena, Murcia, Spain\label{aff8}
\and
European Space Agency/ESTEC, Keplerlaan 1, 2201 AZ Noordwijk, The Netherlands\label{aff9}
\and
Aurora Technology for European Space Agency (ESA), Camino bajo del Castillo, s/n, Urbanizacion Villafranca del Castillo, Villanueva de la Ca\~nada, 28692 Madrid, Spain\label{aff10}
\and
Observatoire Astronomique de Strasbourg (ObAS), Universit\'e de Strasbourg - CNRS, UMR 7550, Strasbourg, France\label{aff11}
\and
Kapteyn Astronomical Institute, University of Groningen, PO Box 800, 9700 AV Groningen, The Netherlands\label{aff12}
\and
Universit\'e Paris-Saclay, Universit\'e Paris Cit\'e, CEA, CNRS, AIM, 91191, Gif-sur-Yvette, France\label{aff13}
\and
Max-Planck-Institut f\"ur Astronomie, K\"onigstuhl 17, 69117 Heidelberg, Germany\label{aff14}
\and
Ohio University, Physics \& Astronomy Department,1 Ohio University, Athens, OH 45701, USA\label{aff15}
\and
NSF's NOIR, Lab 950 N. Cherry Avenue, Tucson, Arizona 85719, USA\label{aff16}
\and
Department of Physics, International University, Ho Chi Minh City, Vietnam\label{aff17}
\and
Vietnam National University, Ho Chi Minh City, Vietnam\label{aff18}
\and
International Space University, 1 rue Jean-Dominique Cassini, 67400 Illkirch-Graffenstaden, France\label{aff19}
\and
School of Physics and Astronomy, University of Leicester, University Road, Leicester, LE1 7RH, UK\label{aff20}
\and
Universit\'e de Franche-Comt\'e, Institut UTINAM, CNRS UMR6213, OSU THETA Franche-Comt\'e-Bourgogne, Observatoire de Besan\c con, BP 1615, 25010 Besan\c con Cedex, France\label{aff21}
\and
INAF-Osservatorio Astrofisico di Torino, Via Osservatorio 20, 10025 Pino Torinese (TO), Italy\label{aff22}
\and
School of Physics, Astronomy and Mathematics, University of Hertfordshire, College Lane, Hatfield AL10 9AB, UK\label{aff23}
\and
Universit\'e Paris-Saclay, CNRS, Institut d'astrophysique spatiale, 91405, Orsay, France\label{aff24}
\and
ESAC/ESA, Camino Bajo del Castillo, s/n., Urb. Villafranca del Castillo, 28692 Villanueva de la Ca\~nada, Madrid, Spain\label{aff25}
\and
INAF-Osservatorio Astronomico di Brera, Via Brera 28, 20122 Milano, Italy\label{aff26}
\and
INAF-Osservatorio di Astrofisica e Scienza dello Spazio di Bologna, Via Piero Gobetti 93/3, 40129 Bologna, Italy\label{aff27}
\and
Dipartimento di Fisica e Astronomia, Universit\`a di Bologna, Via Gobetti 93/2, 40129 Bologna, Italy\label{aff28}
\and
INFN-Sezione di Bologna, Viale Berti Pichat 6/2, 40127 Bologna, Italy\label{aff29}
\and
INAF-Osservatorio Astronomico di Padova, Via dell'Osservatorio 5, 35122 Padova, Italy\label{aff30}
\and
Centre National d'Etudes Spatiales -- Centre spatial de Toulouse, 18 avenue Edouard Belin, 31401 Toulouse Cedex 9, France\label{aff31}
\and
Max Planck Institute for Extraterrestrial Physics, Giessenbachstr. 1, 85748 Garching, Germany\label{aff32}
\and
Universit\"ats-Sternwarte M\"unchen, Fakult\"at f\"ur Physik, Ludwig-Maximilians-Universit\"at M\"unchen, Scheinerstrasse 1, 81679 M\"unchen, Germany\label{aff33}
\and
Dipartimento di Fisica, Universit\`a di Genova, Via Dodecaneso 33, 16146, Genova, Italy\label{aff34}
\and
INFN-Sezione di Genova, Via Dodecaneso 33, 16146, Genova, Italy\label{aff35}
\and
Department of Physics "E. Pancini", University Federico II, Via Cinthia 6, 80126, Napoli, Italy\label{aff36}
\and
INAF-Osservatorio Astronomico di Capodimonte, Via Moiariello 16, 80131 Napoli, Italy\label{aff37}
\and
INFN section of Naples, Via Cinthia 6, 80126, Napoli, Italy\label{aff38}
\and
Instituto de Astrof\'isica e Ci\^encias do Espa\c{c}o, Universidade do Porto, CAUP, Rua das Estrelas, PT4150-762 Porto, Portugal\label{aff39}
\and
Dipartimento di Fisica, Universit\`a degli Studi di Torino, Via P. Giuria 1, 10125 Torino, Italy\label{aff40}
\and
INFN-Sezione di Torino, Via P. Giuria 1, 10125 Torino, Italy\label{aff41}
\and
INAF-IASF Milano, Via Alfonso Corti 12, 20133 Milano, Italy\label{aff42}
\and
Centro de Investigaciones Energ\'eticas, Medioambientales y Tecnol\'ogicas (CIEMAT), Avenida Complutense 40, 28040 Madrid, Spain\label{aff43}
\and
Port d'Informaci\'{o} Cient\'{i}fica, Campus UAB, C. Albareda s/n, 08193 Bellaterra (Barcelona), Spain\label{aff44}
\and
Institute for Theoretical Particle Physics and Cosmology (TTK), RWTH Aachen University, 52056 Aachen, Germany\label{aff45}
\and
INAF-Osservatorio Astronomico di Roma, Via Frascati 33, 00078 Monteporzio Catone, Italy\label{aff46}
\and
Dipartimento di Fisica e Astronomia "Augusto Righi" - Alma Mater Studiorum Universit\`a di Bologna, Viale Berti Pichat 6/2, 40127 Bologna, Italy\label{aff47}
\and
Institute for Astronomy, University of Edinburgh, Royal Observatory, Blackford Hill, Edinburgh EH9 3HJ, UK\label{aff48}
\and
Jodrell Bank Centre for Astrophysics, Department of Physics and Astronomy, University of Manchester, Oxford Road, Manchester M13 9PL, UK\label{aff49}
\and
European Space Agency/ESRIN, Largo Galileo Galilei 1, 00044 Frascati, Roma, Italy\label{aff50}
\and
Universit\'e Claude Bernard Lyon 1, CNRS/IN2P3, IP2I Lyon, UMR 5822, Villeurbanne, F-69100, France\label{aff51}
\and
Institute of Physics, Laboratory of Astrophysics, Ecole Polytechnique F\'ed\'erale de Lausanne (EPFL), Observatoire de Sauverny, 1290 Versoix, Switzerland\label{aff52}
\and
UCB Lyon 1, CNRS/IN2P3, IUF, IP2I Lyon, 4 rue Enrico Fermi, 69622 Villeurbanne, France\label{aff53}
\and
Mullard Space Science Laboratory, University College London, Holmbury St Mary, Dorking, Surrey RH5 6NT, UK\label{aff54}
\and
Departamento de F\'isica, Faculdade de Ci\^encias, Universidade de Lisboa, Edif\'icio C8, Campo Grande, PT1749-016 Lisboa, Portugal\label{aff55}
\and
Instituto de Astrof\'isica e Ci\^encias do Espa\c{c}o, Faculdade de Ci\^encias, Universidade de Lisboa, Campo Grande, 1749-016 Lisboa, Portugal\label{aff56}
\and
Department of Astronomy, University of Geneva, ch. d'Ecogia 16, 1290 Versoix, Switzerland\label{aff57}
\and
INAF-Istituto di Astrofisica e Planetologia Spaziali, via del Fosso del Cavaliere, 100, 00100 Roma, Italy\label{aff58}
\and
INFN-Padova, Via Marzolo 8, 35131 Padova, Italy\label{aff59}
\and
Institut d'Estudis Espacials de Catalunya (IEEC),  Edifici RDIT, Campus UPC, 08860 Castelldefels, Barcelona, Spain\label{aff60}
\and
Institut de Ciencies de l'Espai (IEEC-CSIC), Campus UAB, Carrer de Can Magrans, s/n Cerdanyola del Vall\'es, 08193 Barcelona, Spain\label{aff61}
\and
INAF-Osservatorio Astronomico di Trieste, Via G. B. Tiepolo 11, 34143 Trieste, Italy\label{aff62}
\and
Aix-Marseille Universit\'e, CNRS/IN2P3, CPPM, Marseille, France\label{aff63}
\and
Istituto Nazionale di Fisica Nucleare, Sezione di Bologna, Via Irnerio 46, 40126 Bologna, Italy\label{aff64}
\and
FRACTAL S.L.N.E., calle Tulip\'an 2, Portal 13 1A, 28231, Las Rozas de Madrid, Spain\label{aff65}
\and
Dipartimento di Fisica "Aldo Pontremoli", Universit\`a degli Studi di Milano, Via Celoria 16, 20133 Milano, Italy\label{aff66}
\and
Institute of Theoretical Astrophysics, University of Oslo, P.O. Box 1029 Blindern, 0315 Oslo, Norway\label{aff67}
\and
Leiden Observatory, Leiden University, Einsteinweg 55, 2333 CC Leiden, The Netherlands\label{aff68}
\and
Jet Propulsion Laboratory, California Institute of Technology, 4800 Oak Grove Drive, Pasadena, CA, 91109, USA\label{aff69}
\and
Department of Physics, Lancaster University, Lancaster, LA1 4YB, UK\label{aff70}
\and
Felix Hormuth Engineering, Goethestr. 17, 69181 Leimen, Germany\label{aff71}
\and
Technical University of Denmark, Elektrovej 327, 2800 Kgs. Lyngby, Denmark\label{aff72}
\and
Cosmic Dawn Center (DAWN), Denmark\label{aff73}
\and
Institut d'Astrophysique de Paris, UMR 7095, CNRS, and Sorbonne Universit\'e, 98 bis boulevard Arago, 75014 Paris, France\label{aff74}
\and
NASA Goddard Space Flight Center, Greenbelt, MD 20771, USA\label{aff75}
\and
Department of Physics and Helsinki Institute of Physics, Gustaf H\"allstr\"omin katu 2, 00014 University of Helsinki, Finland\label{aff76}
\and
AIM, CEA, CNRS, Universit\'{e} Paris-Saclay, Universit\'{e} de Paris, 91191 Gif-sur-Yvette, France\label{aff77}
\and
Universit\'e de Gen\`eve, D\'epartement de Physique Th\'eorique and Centre for Astroparticle Physics, 24 quai Ernest-Ansermet, CH-1211 Gen\`eve 4, Switzerland\label{aff78}
\and
Department of Physics, P.O. Box 64, 00014 University of Helsinki, Finland\label{aff79}
\and
Helsinki Institute of Physics, Gustaf H{\"a}llstr{\"o}min katu 2, University of Helsinki, Helsinki, Finland\label{aff80}
\and
Aix-Marseille Universit\'e, CNRS, CNES, LAM, Marseille, France\label{aff81}
\and
NOVA optical infrared instrumentation group at ASTRON, Oude Hoogeveensedijk 4, 7991PD, Dwingeloo, The Netherlands\label{aff82}
\and
INFN-Sezione di Milano, Via Celoria 16, 20133 Milano, Italy\label{aff83}
\and
Universit\"at Bonn, Argelander-Institut f\"ur Astronomie, Auf dem H\"ugel 71, 53121 Bonn, Germany\label{aff84}
\and
Dipartimento di Fisica e Astronomia "Augusto Righi" - Alma Mater Studiorum Universit\`a di Bologna, via Piero Gobetti 93/2, 40129 Bologna, Italy\label{aff85}
\and
Department of Physics, Centre for Extragalactic Astronomy, Durham University, South Road, DH1 3LE, UK\label{aff86}
\and
Universit\'e Paris Cit\'e, CNRS, Astroparticule et Cosmologie, 75013 Paris, France\label{aff87}
\and
University of Applied Sciences and Arts of Northwestern Switzerland, School of Engineering, 5210 Windisch, Switzerland\label{aff88}
\and
Institut d'Astrophysique de Paris, 98bis Boulevard Arago, 75014, Paris, France\label{aff89}
\and
Institut de F\'{i}sica d'Altes Energies (IFAE), The Barcelona Institute of Science and Technology, Campus UAB, 08193 Bellaterra (Barcelona), Spain\label{aff90}
\and
Department of Physics and Astronomy, University of Aarhus, Ny Munkegade 120, DK-8000 Aarhus C, Denmark\label{aff91}
\and
Waterloo Centre for Astrophysics, University of Waterloo, Waterloo, Ontario N2L 3G1, Canada\label{aff92}
\and
Department of Physics and Astronomy, University of Waterloo, Waterloo, Ontario N2L 3G1, Canada\label{aff93}
\and
Perimeter Institute for Theoretical Physics, Waterloo, Ontario N2L 2Y5, Canada\label{aff94}
\and
Space Science Data Center, Italian Space Agency, via del Politecnico snc, 00133 Roma, Italy\label{aff95}
\and
Institute of Space Science, Str. Atomistilor, nr. 409 M\u{a}gurele, Ilfov, 077125, Romania\label{aff96}
\and
Dipartimento di Fisica e Astronomia "G. Galilei", Universit\`a di Padova, Via Marzolo 8, 35131 Padova, Italy\label{aff97}
\and
Departamento de F\'isica, FCFM, Universidad de Chile, Blanco Encalada 2008, Santiago, Chile\label{aff98}
\and
INFN-Sezione di Roma, Piazzale Aldo Moro, 2 - c/o Dipartimento di Fisica, Edificio G. Marconi, 00185 Roma, Italy\label{aff99}
\and
Satlantis, University Science Park, Sede Bld 48940, Leioa-Bilbao, Spain\label{aff100}
\and
Institute of Space Sciences (ICE, CSIC), Campus UAB, Carrer de Can Magrans, s/n, 08193 Barcelona, Spain\label{aff101}
\and
Infrared Processing and Analysis Center, California Institute of Technology, Pasadena, CA 91125, USA\label{aff102}
\and
Instituto de Astrof\'isica e Ci\^encias do Espa\c{c}o, Faculdade de Ci\^encias, Universidade de Lisboa, Tapada da Ajuda, 1349-018 Lisboa, Portugal\label{aff103}
\and
Universidad Polit\'ecnica de Cartagena, Departamento de Electr\'onica y Tecnolog\'ia de Computadoras,  Plaza del Hospital 1, 30202 Cartagena, Spain\label{aff104}
\and
Centre for Information Technology, University of Groningen, P.O. Box 11044, 9700 CA Groningen, The Netherlands\label{aff105}
\and
Institut de Recherche en Astrophysique et Plan\'etologie (IRAP), Universit\'e de Toulouse, CNRS, UPS, CNES, 14 Av. Edouard Belin, 31400 Toulouse, France\label{aff106}
\and
INFN-Bologna, Via Irnerio 46, 40126 Bologna, Italy\label{aff107}
\and
IFPU, Institute for Fundamental Physics of the Universe, via Beirut 2, 34151 Trieste, Italy\label{aff108}
\and
INFN, Sezione di Trieste, Via Valerio 2, 34127 Trieste TS, Italy\label{aff109}
\and
SISSA, International School for Advanced Studies, Via Bonomea 265, 34136 Trieste TS, Italy\label{aff110}
\and
Department of Physics and Astronomy, University of British Columbia, Vancouver, BC V6T 1Z1, Canada\label{aff111}}

% Put your abstract here
%
   \abstract{
  We provide an early assessment of the imaging capabilities of the \Euclid space mission to probe deeply into nearby star-forming regions and associated very young open clusters, and in particular to check to what extent it can shed light on the new-born free-floating planet population. 
 % Four \Euclid pointings %using the regular observing sequence are presented here. In just one shot, \Euclid 
 % were obtained in the Orion giant star-forming complex. This region includes the %Horsehead Nebula 
 % Barnard 30 and 33 dark clouds, the IC\,434 \ion{H}{ii} region, the very young Messier 78, NGC\,2023 and $\sigma$\,Orionis open clusters, and one in the Taurus low-mass star-forming region. 
 This paper focuses on a low-reddening region observed in just one \Euclid pointing where the dust and gas has been cleared out by the hot $\sigma$\,Orionis star.  
 % Astrometric and photometric data in four passbands (\IE, \YE, \JE, and \HE ) for over 10$^5$ sources have been extracted in the $\sigma$\,Orionis region, demonstrating the great potential of the \Euclid mission to %
  %one  ROS reaches enough 
  %the substellar-mass population  over 
%  probe unprecedented large areas in nearby star-forming regions. 
One late-M and six known spectroscopically confirmed L-type substellar members in the $\sigma$\,Orionis cluster are used as benchmarks to provide a high-purity procedure to select new candidate members with \Euclid. %Two of these benchmarks appear to be almost resolved binaries or multiple systems.  
  The exquisite angular resolution and depth delivered by the \Euclid instruments allow us to focus on bona-fide point sources. A cleaned sample of $\sigma$\,Orionis cluster substellar members has been produced and the initial mass function (IMF) has been estimated by combining \Euclid and \textit{Gaia} data. Our  $\sigma$\,Orionis substellar IMF is consistent with a power-law distribution with no significant steepening at the planetary-mass end. No evidence of a low-mass cutoff is found down to about 4 Jupiter masses at the young age (3 Myr) of the $\sigma$\,Orionis open cluster.
  %, although it seems that the IMF slope is shallower inside the planetary-mass domain. 
}
%
% Provide up to five key words:
%

 \titlerunning{\Euclid: ERO -- Free-floating new-born planets in the $\sigma$\,Orionis cluster}
   \authorrunning{E. L. Martín et al.}
   
   \maketitle
%
%-------------------------------------------------------------------
%
%
%   Start the main text of your paper here
%
   
\section{\label{sc:Intro}Introduction}
The nearest star-forming regions provide us with a natural laboratory to investigate in detail the complex processes that transform molecular clouds into stellar and substellar-mass objects. In particular, one of the long-standing questions is whether or not there is a low-mass cutoff in the initial mass function (IMF). The original IMF was defined by \citet{Salpeter55} as a single power-law function over the mass range from 10 down to 0.4 Solar masses (M$_\odot$). While the early computations of spherical collapse including dust-grain opacities found a minimum mass of 0.1\,M$_\odot$ owing to opacity-limited fragmentation, i.e., above the substellar-mass limit \citep{Silk77}, recent calculations predict that the minimum fragment could reach down to 10$^{-3}$\,M$_\odot$ \citep{Mondal19}, i.e. well below the deuterium-burning mass limit. The thermonuclear fusion of deuterium, $^2$H(p,$\gamma$)$^3$He, takes place at 10$^6$\,K and can be important in the early stages of evolution of objects with masses above 13 times the mass of Jupiter (1\,M$_{\rm J}$ = 0.000955 M$_\odot$); see  \citet{Bodenheimer66}, \citet{Stahler88}, and \citet{2000Chabrier}.   

Deep observations of stellar nurseries, very young open clusters, and young stellar associations have been made to search for the predicted low-mass cutoff of the IMF, and they have reported that the IMF extends smoothly into the realm of planetary masses, reaching down to the deuterium limit and  overlapping with the masses of exoplanets. Various names have been used to refer to these unexpected substellar-mass objects, such as, for example, brown dwarfs (BDs) of planetary mass, sub-brown dwarfs, cluster planets, nomadic worlds, free-floating planets (FFPs), rogue planets, and planetary-mass objects (PMOs). Collectively, substellar-mass objects are ultracool dwarfs (UCDs) with very cool effective temperatures, late spectral types, small sizes and faint luminosities that make them appear to be a tiny minority among the myriad of stars and galaxies in deep astronomical surveys, even though their numbers can be significant.  

Free-floating planets appear to be ubiquitous and numerous, since they have been identified by direct imaging and spectroscopy in many different stellar cradles. Some examples of such targets are: the Chamaeleon I star-forming region \citep{1999Oasa, 2004Luhman}; the IC\,348 and NGC\,1333 clusters in Perseus \citep{2017Esplin, Scholz2023}; the Ophiucus star-forming region \citep{2015Chiang, Bouy2022}; the Orion Nebula cluster \citep{Lucas2000, Lucas2001, 2006Lucas}; the  Lynds 1630 molecular clouds \citep{ 2015Spezzi}, the $\sigma$\,Orionis \citep{Zapatero2000, 2009Lodieu} and Collinder 69 \citep{2011Bayo} young clusters in the Orion giant star-formation complex; the Upper Sco OB association \citep{Lodieu2018, Lodieu2021, Miret-Roig2022}; and the Taurus dark clouds \citep{2019Esplin}. PMOs have also been found as wide companions to stars and BDs \citep{2005Chauvin, 2015Gauza}, as members of young moving associations \citep{2021Zhang}, and as microlensing events towards the Galactic bulge \citep{2018Mroz, 2023Koshimoto, 2023Sumi}.    

The existence of FFPs challenges models of star and planet formation. A variety of physical mechanisms have been proposed to explain the formation of substellar objects with masses well below the Jeans limit, the leading one being turbulent fragmentation \citep{2004Padoan, 2008Hennebelle}, but others, including gravitational collapse in filaments, ejection from proto-planetary discs, and photo-erosion \citep{2023Miret} have not been discarded as potential players.  

The cosmologically-driven requirements of the \Euclid mission \citep{Laureijs11} and the performance of its VIS \citep{EuclidSkyVIS} and NISP \citep{EuclidSkyNISP} instruments are expected to enable a major leap in sensitivity gain and area coverage that will foster the advance of many areas of legacy science in astrophysics \citep{EuclidSkyOverview}, including the detection of around a million UCDs over a large portion of the Milky Way \citep {2021Solano,2023Martin}, with spectroscopic reconnaissance spectra for thousands of them \citep {2021Martin,2024Jzhang}. %such as SDSS \citep{sdss}, Gaia \citep{gaia}, and DES \citep{des}
The \Euclid reference observing sequence (ROS) is the main observation mode that is used for the wide and deep surveys. It is required to reach limiting AB magnitudes of $26.2$ in the optical $\IE$ band and of $24.5$ in the near-infrared NISP bands over a wide area \citep{Scaramella-EP1}. 
%EROData

The \Euclid Early Release Observations (ERO) programme has been designed to be a showcase of the mission´s potential for legacy science across a wide range of sky regions. 
It demonstrates that \Euclid brings a unique combination of unprecedented sensitivity, wide-area coverage, and high spatial resolution to the investigation of diverse science topics. The first ERO papers include studies of very high-redshift objects \citep{EROLensVISDropouts}, clusters of galaxies \citep{EROPerseusICL,EROPerseusDGs,EROFornaxGCs}, nearby galaxies \citep{ERONearbyGals}, and galactic globular clusters \citep{EROGalGCs}.  
 
This ERO paper investigates the power of \Euclid to probe deep into very young regions over a wide area, reaching detection limits capable of revealing the FFP population and eventually search for the predicted low-mass cutoff of the IMF. The paper is structured as follows. In Sect.~\ref{sc:Observation}, the general \Euclid ERO project number~2 (ERO02, P.I. Martín) is presented. Five \Euclid pointings were obtained and this work focuses in about half of the area covered by one of them. In Sect.~\ref{sc:Results}, the particular region that is the focus of this work is described and in Sect.~\ref{sc:SOri} previously known substellar-mass objects in the $\sigma$\,Orionis cluster are discussed and the cuts used to select new FFP candidates are described. %Section~\ref{sc:Pairs} presents a search for wide companions around substellar primaries that yields several candidates. 
Section~\ref{sc:IMF} deals with the revised initial mass function (IMF) of the $\sigma$\,Orionis cluster in the area covered by the \Euclid observations and comparison with the field IMF low-mass tail.  
Finally, Sect.~\ref{sc:final}
summarises our results and provides future prospects.

\begin{figure}[htbp!]
\centering
\includegraphics[angle=0,width=\linewidth]{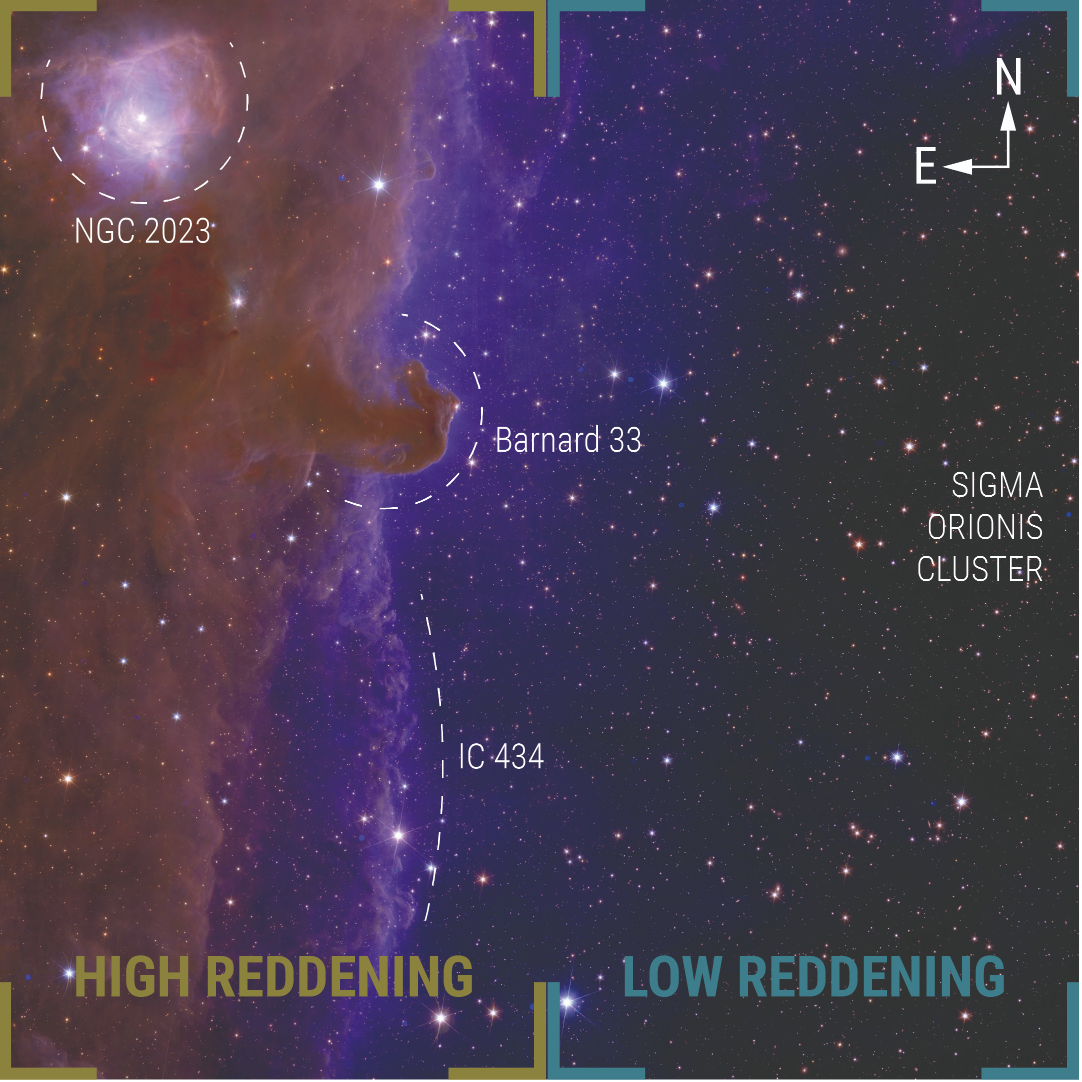}
\caption{Multi-colour mosaic of the {\it Euclid\/}  pointing studied in this work. The area covered is 0.58 square degrees. The dark neck of the HorseHeard (Barnard 33) is pointing towards the bright $\sigma$\,Orionis star, located just outside the field of view (FoV). The bright nebular emission crossing the image is the IC\,434 \ion{H}{ii} region, and the bright concentration at the upper left corner is NGC\,2023. This paper focuses on the low reddening part of the FoV that has been cleared out by the hot $\sigma$\,Orionis star. }
\label{fig:mosaic}
\end{figure}

\section{The \Euclid Early Release Observations project of nearby star-forming regions %of Barnard 33 and the is cluster
\label{sc:Observation} }

This \Euclid ERO programme has targeted nearby (distance $\le 400\,$pc) star-forming regions and very young open clusters (age $<10$\,Myr) to explore their faint ultra-cool populations, search for FFPs, and determine whether or not there is an IMF low-mass cutoff. The total project consists of five \Euclid pointings. 
The targets were the following: the NGC\,1333 cluster in Perseus (incomplete dataset); the Barnard 30, Barnard 33 (Horsehead nebula, that also includes the NGC\,2023 embedded cluster and part of the $\sigma$\,Orionis open cluster), and Messier 78 dark clouds in the Orion star-formation complex, and a field containing several dark clouds in the Taurus region. In this paper we focus on one of these targets, called the Horsehead field, and in particular we focus on about half of the area, hereafter nicknamed the ERO-SOri field. The other regions covered by the \Euclid ERO pointings will be the subject of future studies.  

\def\degf {\hbox{$.\!\!^{\circ}$}}

The \Euclid observation of the ERO-SOri field took place on 2 October 2023. A full ROS with good guiding was obtained.
  The centre coordinates of each of the four \Euclid exposures that make up the ROS were the following: 
  85\degf150915,     $-$2\degf613342; 
  85\degf167068,    $-$2\degf582078; 
  85\degf166265,    $-$2\degf551255; and 
  85\degf182417,    $-$2\degf519991. The full FoV of the ERO pointing presented here is displayed in Fig.~\ref{fig:mosaic} and covers an area of 0.58 square degrees. The FoV was chosen to avoid the blinding star $\sigma$\,Orionis and to include the Barnard 33 molecular cloud (Horsehead Nebula), the NGC\,2023 cluster and reflection nebula, and the IC\,434 \ion{H}{ii} region. The full \Euclid ROS consisted of four dithered exposures in VIS and NISP using the nominal exposure times described in \cite{EuclidSkyVIS} and \cite{EuclidSkyNISP}, respectively. %of 89.5\,{\rm s} and a long exposure time of 560.5\,{\rm s}. Simultaneously, four dithered exposures were also obtained with the NISP instrument, each of them with an exposure time of 87.2\,{\rm s} in each of the three photometric passbands and 549.6\,{\rm s} in the slitless spectroscopic mode. 
  The dithering pattern is designed so that the gaps between the detectors can be covered when making a stack of the four images. However, due to a failure in the implementation of the dithering during the science-verification phase, the pattern was not optimised and there are some gaps in the mosaic of this ERO footprint. Furthermore, 
during data processing it was realised that about 5\% of the FoV was covered by only one image and cosmic rays could not be removed efficiently. 
After data reduction and image stacking, following the procedures described in \cite{EROData}, the data have been validated and considered ready for science exploitation. In this work, the catalogues and images of the ERO public data release are used  \citep{EROcite}. They do not include any spectroscopic data.

\begin{figure}[htbp!]
\centering
\includegraphics[width=\linewidth]{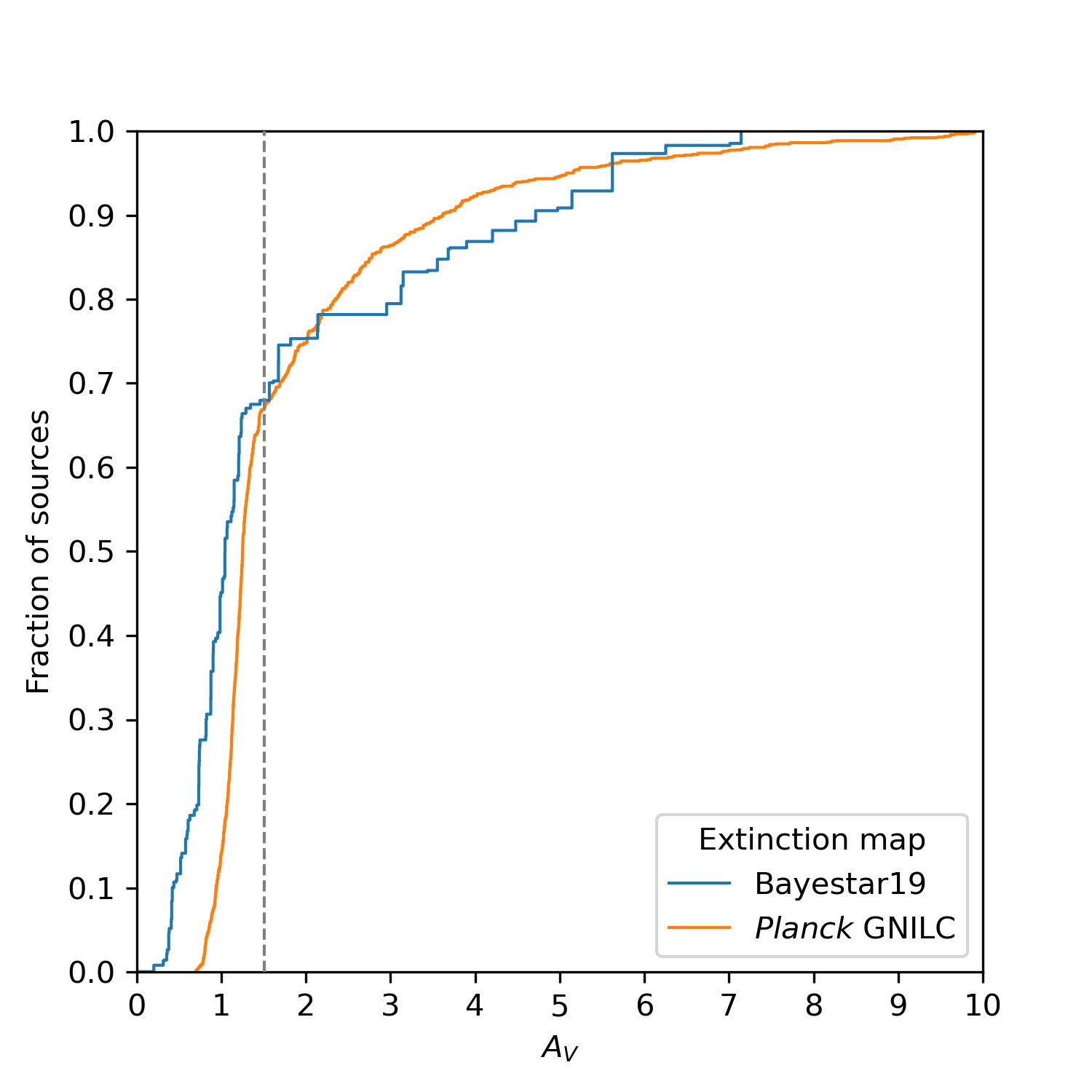}
\caption{Cumulative distribution of interstellar reddening in the {\it Euclid\/} ERO region shown in Fig.~\ref{fig:mosaic}. Two different methods of estimating the reddening are compared. 
Note that about 70\% of the FoV has a modest reddening of less than 1.7 mag in the visual. This reddening is used in the reddening vectors shown in the colour-magnitude and u-colour diagrams. This work focuses on the low-reddening region of the FoV.}
\label{fig:redd}
\end{figure}

\section{The region covered in the ERO observation 
%results in the Barnard 33, IC\,434 and the $\sigma$Orionis cluster 
\label{sc:Results} }

The ERO pointing shown in Fig.~\ref{fig:mosaic} contains the complex region created by the interaction between the hot $\sigma$\,Orionis star and the Orion B giant molecular cloud (Lynds 1630). 
Extreme ultraviolet radiation from the O-type star $\sigma$\,Orionis creates a bright ionisation front that is known as the IC\,434 \ion{H}{ii} region. A complicated pattern of bright and dark regions is clearly seen in fine detail in the \Euclid mosaic (Fig.~\ref{fig:mosaic}). The Horsehead Nebula (Barnard 33) is projected in the foreground of the \ion{H}{ii} region at a distance of around 360\,pc and it points towards the ionising $\sigma$\,Orionis star that is in the background \citep{2018Bally}. 
Another source of ionisation in the ERO-SOri FoV is the B-type star HD\,37903 that illuminates a reflection nebula and is associated with the embedded open cluster NGC\,2023 that contains very young low-mass stars  \citep{1990Depoy,2009Mookerjea,2017kounkel}. 
As a consequence of the complex past and ongoing star-formation processes, there are patches with significant interstellar reddening. 
The $A_V$ extinction values for all the sources identified in this ERO pointing have been calculated with two extinction maps from the literature: the generalised needlet internal linear combination map from \citet{2016A&A...596A.109P}, which is a 2D extinction map, and Bayestar19 \citep{2018Green,2019ApJ...887...93G}, which is a 3D extinction map for which we assumed a mean distance of 400\,pc to the Orion star-forming region. Both extinction maps were queried with the \texttt{dustmap}\footnote{\url{https://dustmaps.readthedocs.io/en/latest/index.html}} package. The cumulative distribution function of the $A_V$ extinction of our sources is shown in Fig.~\ref{fig:redd}, 
   where the curves depict the distributions of the two extinction maps from the literature and the vertical dashed line shows the maximum extinction value that our selection criteria can cover. As can be observed, both extinction maps agree quite well and show that about 70\% of the sources in the FoV have $A_V$ values between 0 and 1.7 mag. In the future we plan to use more specific methods that take into account the extinction of the individual sources \citep[e.g.,][]{2021A&A...649A.159O}. 

\section{The \Euclid view of the \texorpdfstring{$\sigma$\,Orionis}{sigma Orionis} substellar members 
\label{sc:SOri} }

The \Euclid footprint of the ERO pointing includes a portion of the well-known $\sigma$\,Orionis cluster that has been a favourite hunting ground for very young substellar objects and FFPs for over two decades \citep{Zapatero2000,2023Damian}. A review of the $\sigma$\,Orionis cluster properties was provided by \citet{2008Walter}. A recent assessment of cluster membership using the \textit{Gaia} third data release \citep[DR3, ][]{2023A&A...674A...1G} has been carried out in a study of the young populations in the region \citep{2023Zerjal}. 
The ages of most $\sigma$\,Orionis cluster members are in the range 1--5\,Myr \citep{2002Zapatero, 2023Zerjal}. In this work an age of $(3\pm2)$\,Myr and a distance of 402.74$\pm$9\,pc are adopted for the $\sigma$\,Orionis cluster. 
The deepest survey carried out to date in the search for FFPs belonging to $\sigma$\,Orionis has been reported by \cite{2012Ramirez} using ground-based telescopes.

\subsection{Definition of benchmarks for \Euclid based on confirmed \texorpdfstring{$\sigma$\,Orionis}{sigma Orionis} substellar objects 
\label{sc:SOri1} }

%They used  and from the Canada-France-Hawaii Telescope/WIRCam survey by \cite{2023Damian}
We selected seven confirmed substellar-mass members of the $\sigma$\,Orionis cluster with ground-based low-resolution optical and near-infrared spectroscopic classification. Their names and coordinates are listed in Table~\ref{tab:table1}, together with the spectral types from the literature \citep{Zapatero2000,Barrado2001,Martin2001,2017Zapatero}. The parameters of these seven benchmarks in the \Euclid ERO catalogue are listed in Table~\ref{tab:table2}. 

\begin{table}[htbp!]  \caption{Benchmark $\sigma$\,Orionis cluster members observed with \Euclid.}
\setlength{\tabcolsep}{5pt} %6pt is standard
\begin{tabular}{lccc}
\hline\hline
\noalign{\vskip 1pt}
%Brief name        & RA & Dec & \begin{tabular}[c]{@{}l@{}}Sp.T.\end{tabular} \\ \hline
Nickname        & RA & Dec & Spectral \\
            & [hh mm ss.ss] & [deg mm ss.s] &      type     \cr
\noalign{\vskip 1pt}
 \hline\noalign{\vskip 1pt}
S\,Ori\,28             & 05 39 23.19 & $-$02 46 55.8  & M5.5     \\
\hline\noalign{\vskip 1pt}
S\,Ori\,52             & 05 40 09.20 & $-$02 26 32.0  & L0.5     \\
\hline\noalign{\vskip 1pt}
S\,Ori\,60             & 05 39 37.50 & $-$02 30 42.0  & L2.0     \\
\hline\noalign{\vskip 1pt}
S\,Ori\,62             & 05 39 42.05 & $-$02 30 31.6  & L4.0     \\
\hline
S\,Ori\,054017 & 05 40 17.34 & $-$02 36 22.6  & L3.5      \\
\hline\noalign{\vskip 1pt}
S\,Ori\,054000 & 05 40 00.04 & $-$02 40 33.1  & L2.0     \\
 \hline\noalign{\vskip 1pt}
S\,Ori\,054037 & 05 40 37.82 & $-$02 40 01.1  & L4.5    \\
 \hline
\end{tabular}
\label{tab:table1}
\end{table}

\begin{figure*}[htbp!]
\centering
\includegraphics[width=\linewidth]{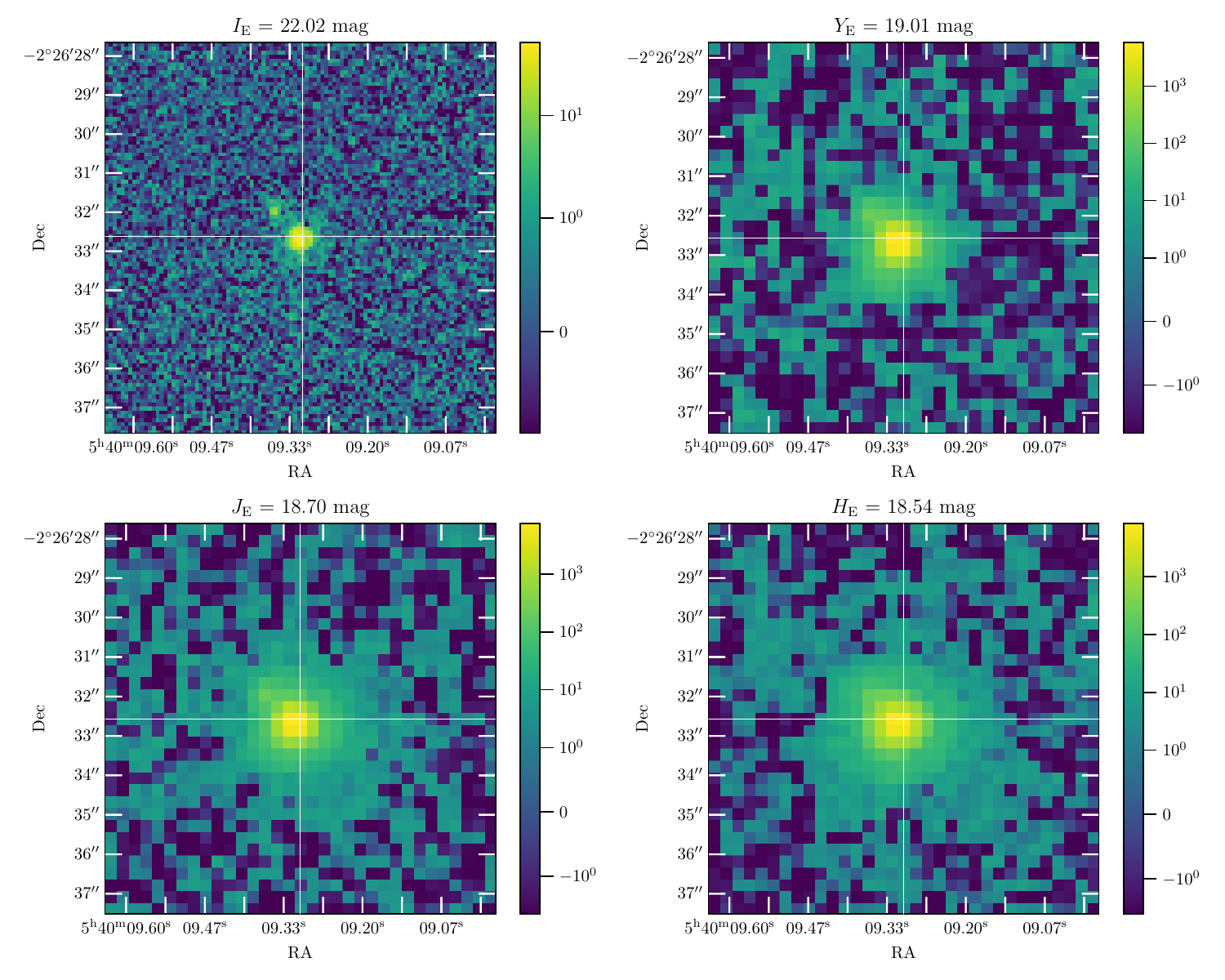}
\caption{Mosaic of {\it Euclid} images centred on the benchmark object S\,Ori\,52 in the four different photometric passbands. We note the presence of a clearly resolved visual companion in the VIS image.}
\label{fig:SOri52}
\end{figure*}

\begin{table*}[htbp!]
 \caption{\Euclid parameters for $\sigma$\,Orionis benchmark objects. FWHM values are given in pixels. }
 \setlength{\tabcolsep}{4pt} %6pt is standard
%\smallskip
%\label{table:obslog}
%\smallskip
%\small
\begin{tabular}{lccccccc}
 \hline\hline\noalign{\vskip 1pt}
  & \multicolumn{7}{c}{Object} \\
\cline{2-8}\noalign{\vskip 1pt}
{Parameter} & S   Ori 28 & S\,Ori\,52 & S   Ori 60 & S   Ori 62 & S\,Ori\,054017 & S\,Ori\,054000 & S\,Ori\,054037 \\
\hline\noalign{\vskip 1pt}
SPREAD\_MODEL\_I & 1.20$\times 10^{-3}$   & $-$7.20$\times 10^{-4}$  & 3.14$\times 10^{-4}$   & $-$7.06$\times 10^{-4}$  & 2.94$\times 10^{-3}$       & 1.82$\times 10^{-3}$        & 1.40$\times 10^{-3}$          \\ \hline\noalign{\vskip 1pt}
SPREAD\_MODEL\_Y & 1.24$\times 10^{-2}$   & 6.25$\times 10^{-3}$   & 6.96$\times 10^{-6}$   & 6.61$\times 10^{-3}$   & 7.24$\times 10^{-3}$        & 1.48$\times 10^{-4}$        & $-$2.75$\times 10^{-4}$         \\ \hline\noalign{\vskip 1pt}
SPREAD\_MODEL\_J & 9.24$\times 10^{-3}$   & 3.37$\times 10^{-4}$   & $-$1.59$\times 10^{-3}$  & $-$2.14$\times 10^{-3}$  & $-$5.83$\times 10^{-3}$       & $-$1.59$\times 10^{-4}$       & $-$3.76$\times 10^{-3}$         \\ \hline\noalign{\vskip 1pt}
SPREAD\_MODEL\_H & 6.80$\times 10^{-3}$   & 3.26$\times 10^{-3}$   & 3.74$\times 10^{-3}$   & 5.47$\times 10^{-3}$   & $-$1.23$\times 10^{-3}$       & $-$4.61$\times 10^{-3}$       & 1.10$\times 10^{-2}$          \\ \hline\noalign{\vskip 1pt}
CLASS\_STAR\_I   & 0.98       & 1          & 0.81       & 0.86       & 0.68            & 0.76            & 0.75              \\ \hline\noalign{\vskip 1pt}
CLASS\_STAR\_Y   & 0.99       & 1          & 1          & 0.92       & 0.98            & 0.98            & 1                 \\ \hline\noalign{\vskip 1pt}
CLASS\_STAR\_J   & 1          & 1          & 1          & 0.98       & 0.98            & 0.98            & 0.98              \\ \hline\noalign{\vskip 1pt}
CLASS\_STAR\_H   & 1          & 1          & 1          & 0.94       & 0.98            & 0.98            & 0.97              \\ \hline\noalign{\vskip 1pt}
FWHM\_IMAGE\_I   & 1.69       & 2.46       & 2.39       & 1.51       & 1.73            & 1.69            & 1.65              \\ \hline\noalign{\vskip 1pt}
FWHM\_IMAGE\_Y   & 1.44       & 1.52       & 0.78       & 1.78       & 1.61            & 1.59            & 1.42              \\ \hline\noalign{\vskip 1pt}
FWHM\_IMAGE\_J   & 1.83       & 1.54       & 0.8        & 1.46       & 1.29            & 1.66            & 1.37              \\ \hline\noalign{\vskip 1pt}
FWHM\_IMAGE\_H   & 1.79       & 1.56       & 1.69       & 1.89       & 1.64            & 1.4             & 1.56              \\ \hline\noalign{\vskip 1pt}
MAG\_AUTO\_I     & 18.4631    & 21.9735    & 23.7263    & 23.8577    & 24.9944         & 24.2328         & 24.3397           \\ \hline\noalign{\vskip 1pt}
MAGERR\_AUTO\_I  & 0.0005     & 0.0048     & 0.023      & 0.0206     & 0.0537          & 0.0268          & 0.0305            \\ \hline\noalign{\vskip 1pt}
MAG\_AUTO\_Y     & 16.3237    & 18.8565    & 20.597     & 20.5435    & 21.5509         & 20.8739         & 21.0935           \\ \hline\noalign{\vskip 1pt}
MAGERR\_AUTO\_Y  & 0.001      & 0.0035     & 0.012      & 0.0118     & 0.0305          & 0.0215          & 0.0193            \\ \hline\noalign{\vskip 1pt}
MAG\_AUTO\_J     & 16.3058    & 18.5677    & 20.1688    & 20.1503    & 21.0523         & 20.4058         & 20.6629           \\ \hline\noalign{\vskip 1pt}
MAGERR\_AUTO\_J  & 0.0008     & 0.0023     & 0.0068     & 0.007      & 0.0162          & 0.0119          & 0.0111            \\ \hline\noalign{\vskip 1pt}
MAG\_AUTO\_H     & 16.3073    & 18.3369    & 19.7774    & 19.8562    & 20.5806         & 20.0368         & 20.1653           \\ \hline\noalign{\vskip 1pt}
MAGERR\_AUTO\_H  & 0.0008     & 0.002      & 0.0051     & 0.0055     & 0.0121          & 0.0089          & 0.0075            \\ \hline\noalign{\vskip 1pt}
ELLIPTICITY\_I  & 0.048     & 0.100      & 0.243     & 0.049     & 0.045          & 0.106          & 0.028            \\ \hline\noalign{\vskip 1pt}
ELLIPTICITY\_J  & 0.017     & 0.049      & 0.023     & 0.04     & 0.07          & 0.107          & 0.062            \\ \hline
\end{tabular}
%\normalsize
\label{tab:table2}
\end{table*}

The values of the SPREAD$_-$MODEL parameter for the benchmarks in all the \Euclid passbands have very small deviations from zero, as expected for bona-fide point sources. This parameter was developed as a star/galaxy classifier by the data management pipeline of the Dark Energy Survey \citep{2012Mohr}, and has been shown to be a good discriminant for point sources in nearby young clusters and stellar associations \citep{2013Bouy}. The parameter is adopted in this work as one of the main selection criteria to separate point sources from galaxies.   

We note that two benchmarks, namely S\,Ori\,52 and 60, have FWHM$_-$IMAGE$_-$I values in the ERO catalogue that are slightly larger than the other four benchmarks, suggesting that they might be binaries that have angular separations close to the limit of spatial resolution of the VIS data. Moreover, a resolved faint visual companion was spotted close to S\,Ori\,52 at position J054009.36$-$022631.93 in the VIS image (see Fig.~\ref{fig:SOri52}). S\,Ori\,52 has an optical spectral type of L0.5 and a mass around $15\,M_{\rm J}$ \citep{2001Bejar}. 
The candidate wide companion to S\,Ori\,52 is 3.2 mag fainter in the $\IE$ passband than the primary, the angular separation is \ang{;;0.962} (387\,au at 402\,pc) and the position angle is \ang{43.3;;}. The pair is not fully resolved in the NISP images because they have lower spatial resolution than the VIS image. Using PSF photometry, the difference in magnitude in the 
$\JE$ passband is 3.96. This difference is larger than in 
$\IE$, indicating that the companion has slightly bluer $\IE - \JE$ colour than the primary and casting some doubt on the physical association of these two objects. 
The possibility that \Euclid may have found two substellar binaries, close to the angular resolution of the VIS images, in a sample of only seven benchmarks in the $\sigma$\,Orionis cluster, is interesting and deserves further scrutiny. A {\it Hubble} Space Telescope (HST) imaging survey of wide binaries (projected semi-major axes between 100 and 1000 au) among pre-main-sequence (PMS) stars in the Orion star-forming complex found a binary frequency of 12.5$_{-0.8}^{+1.2}$\% \citep{2016Kounkel}, and recent work with the {\it James Webb} Space Telescope suggests that substellar-mass binaries in the Trapezium cluster could be common \citep{McCaughrean}. On the other hand, 
no resolved binaries with separations $>\rm{20\,au}$ were found in an imaging survey of 33 BDs in two young open clusters (ages in the range from 70 to 120\,Myr) carried out with the HST \citep{2003Martin}.

\subsection{Contamination estimates and definition of selection cuts for the \Euclid data
\label{sc:contamination} }

To assess the likelihood of contamination of substellar object candidates by background extragalactic objects and by foreground ultracool dwarfs, all the objects not saturated in the \Euclid images and listed by \cite{2012Ramirez} in the ERO-SOri FoV were visually inspected in the VIS and NISP images and were cross-correlated with the ESO VISTA Hemisphere Survey (VHS) catalogue \citep{2021McMahon} to check for proper motions. 
The total proper motion of true cluster members is expected to be $\le20\,{\rm mas}$ per year, and thus should not be measurable when comparing VHS and NISP data. A summary of the results of this contamination assessment is provided in Table~\ref{tab:table3}. Sources that are spatially resolved as extended objects in any of the \Euclid passbands are considered as non-members. They have values of the FWHM and SPREAD$_-$MODEL parameters larger than those of the benchmarks. Sources that are detected to move by $\ge100\,{\rm mas}$ from the VHS epoch to the \Euclid epoch (baseline 14 years) are classified as non-members and are labelled as high proper motion. For the benchmarks, we checked that their coordinates match within $100\,{\rm mas}$ with those from the VHS catalogue. As expected, the most frequent contamination comes from extended objects that are probably background galaxies (9/38 = 24\,\%), particularly at the faint end of the sample. The ratio of extragalactic sources to ultracool dwarfs is expected to get larger with increasing depth. It has recently been reported from JWST/NIRSpec spectroscopic follow-up of photometrically selected JWST/NIRcam compact sources that the ratio between extragalactic objects and ultracool dwarfs is 11/3 at depths fainter than those reached by  the \Euclid images \citep{2023Langeroodi}.

\begin{table*}[htbp!]  \caption{Previously identified candidate members of the $\sigma$\,Orionis cluster not validated with \Euclid data.}
\smallskip
\label{table:obslog}
\smallskip
%\small
\begin{tabular}{lccl}
  \hline\hline\noalign{\vskip 1pt}
Name & RA & Dec 
%& $\JE$, err & $\HE$, err 
& Comment \\
            & [hh mm ss.ss] & [deg mm ss.s] &           \cr
 \hline\noalign{\vskip 1pt}
S\,Ori\,69  & 05 39 18.05 & $-$02 28 54.1  & extended \\
S\,Ori\,J053923$-$021235 & 05 39 23.28 & $-$02 12 35.0 & extended \\
%S\,Ori\,42  & 05 39 23.43 & $-$02 40 57.5 & 17.583, 0.002 & 17.490, 0.002 & member \\
S\,Ori\,J053929$-$024636 & 05 39 29.36 & $-$02 46 37.1 & high proper motion \\
%S\,Ori\,J053932$-$025220 & 05 39 32.43 & $-$02 52 20.2 & 20.547, 0.014 & 20.066, 0.011 & member \\
%S\,Ori\,60  & 05 39 37.50 & $-$02 30 42.0 & 20.175, 0.012 & 19.766, 0.009 & member \\
%S\,Ori\,62  & 05 39 42.05 & $-$02 30 31.6 & 20.144, 0.008 & 19.840, 0.006 & member \\
%S\,Ori\,J053946$-$022423 & 05 39 46.46 & $-$02 24 23.2 & 18.083, 0.002 & 17.948, 0.003 & member \\
S\,Ori\,57  & 05 39 47.05 & $-$02 25 24.5 & high proper motion \\
%S\,Ori\,J053949$-$023130 & 05 39 49.52 & $-$02 31 29.9  & 19.932, 0.008 & 19.588, 0.006 &  member \\
%S\,Ori\,J053953$-$021622  & 05 39 53.06 & $-$02 16 22.9  & 17.811, 0.003 & 17.670, 0.003 & member \\
S\,Ori\,J053956$-$025315  & 05 39 56.81 & $-$02 53 14.6  %& 19.350, 0.007 & 
& extended \\
%S\,Ori\,J053957$-$025006  & 05 39 57.40 & $-$02 50 06.2  & 20.013, 0.010 & 19.700, 0.009 & member \\
%S\,Ori\,J054000$-$024033 & 05 40 00.04 &  $-$02 40 33.1 & 20.411, 0.010 & 20.026, 0.010 & member \\
S\,Ori\,J054004$-$025332 & 05 40 04.48 & $-$02 53 31.9 & extended \\ 
%S\,Ori\,J054006$-$023605 & 05 40 06.95 &  $-$02 36 05.1 & 19.154, 0.006 & 18.865, 0.005 & member \\
%S\,Ori\,J054007$-$022234 & 05 40 07.75 &  $-$02 22 34.3 & 20.251, 0.011 & 19.881, 0.009 & member \\
%S\,Ori\,J054008$-$024551 & 05 40 08.51 & $-$02 45 50.6 & 20,185, 0.009 & 19.726, 0.007 & member \\
%S\,Ori\,52  & 05 40 09.20 & $-$02 26 32.0 & 18.560, 0.003 & 18.306, 0.003 & member \\
PBZ12 J054011$-$025639 & 05 40 11.62 & $-$02 56 39.4 & high proper motion \\  
S\,Ori\,J054014$-$025146  & 05 40 14.23 & $-$02 51 46.3 &  extended \\
%S\,Ori\,J054017-023623 & 05 40 17.34 & $-$02 36 22.6 & 21.030, 0.018 & 20.455, 0.013 & member \\
%Gaia DR3 3216520111991686784  & 05 40 18.38 & $-$02 23 08.8 & 17.184, 0.002 & 16.468, 0.002 & member \\
%TIC 11360257 & 05 40 20.79 & $-$02 24 00.1  & 18.071, 0.003 & 17.925, 0.003 & member \\
PBZ12 J054024$-$024444 & 05 40 24.32 & $-$02 44 44.3  & extended \\
PBZ12 J054025$-$024259 & 05 40 25.39 & $-$02 42 59.7  & extended \\
PBZ12 J054026$-$023100 & 05 40 26.44 & $-$02 31 00.8 &  high proper motion \\ 
PBZ12 J054028$-$025116 & 05 40 27.99 & $-$02 51 16.7 & extended \\
PBZ12 J054038$-$022806 & 05 40 38.42 & $-$02 28 06.6 & extended \\
%PBZ12 J054040$-$023810 & 05 40 40.73 & $-$02 38 10.8 & 21.561, 0.027 & 21.207, 0.020 & member  \\
\hline
\end{tabular}
\label{tab:table3}
\end{table*}

%New Table added
\begin{table*}[htbp!]
\caption{ \Euclid point-source selection criteria. The right ascension, RA, divides the two area into parts: the high-reddening region (${\rm RA} < 85\degf1875$); and the low reddening region (${\rm RA}\ge  85\degf1875)$.}
\begin{tabular}{llcc@{\hskip 1.5em}cc}
\hline\hline\noalign{\vskip 1pt}
 & & \multicolumn{2}{c}{High-reddening region} & \multicolumn{2}{c}{Low-reddening region}           \\
\cline{3-6}\noalign{\vskip 1pt}
Cut No. & Criteria & Count after cut & \%  of catalogue & Count after cut & \% of catalogue \\ \hline\noalign{\vskip 1pt}
1  & ra\_J\_E Cut & 124\,249 & 38.04 & 126\,693 & 38.79 \\
\hline\noalign{\vskip 1pt}
2  & $-$0.011\textless{}SPREAD\_MODEL\_J\_E\textless{}0.014 &70\,558 & 21.60 & 47\,630 & 14.58 \\
\hline\noalign{\vskip 1pt}
3  & 0.472\textless{}FWHM\_IMAGE\_J\_E\textless{}2.158 & 24\,174       & 7.40 & 14\,932 & 4.57 \\
\hline\noalign{\vskip 1pt}
4  & 0\textless{}ELLIPTICITY\_J\_E\textless{}0.138 & 17\,423 & 5.33 & 10\,742 & 3.29         \\
\hline\noalign{\vskip 1pt}
5  & $-$0.01\textless{}SPREAD\_MODEL\_H\_E\textless{}0.016 & 14\,574 & 4.81 & 7944 & 2.97 \\
\hline\noalign{\vskip 1pt}
6  & 1.238\textless{}FWHM\_IMAGE\_H\_E\textless{}2.052 & 12\,262 & 3.87 & 7018 & 2.50 \\
\hline\noalign{\vskip 1pt}
7  & $-$0.005\textless{}SPREAD\_MODEL\_Y\_E\textless{}0.017 &11\,550 & 3.62 & 6603 & 2.30 \\
\hline\noalign{\vskip 1pt}
8  & 0.461\textless{}FWHM\_IMAGE\_Y\_E\textless{}2.099 & 10\,768 & 3.34 & 6208 & 2.03 \\
\hline\noalign{\vskip 1pt}
9  & Drop 99 in any I\_E mag & 9450 & 2.89 & 4663 & 1.43 \\
\hline\noalign{\vskip 1pt}
10 & $-$0.002\textless{}SPREAD\_MODEL\_I\_E\textless{}0.004 &6819 & 2.09 & 3445 & 1.05 \\
\hline\noalign{\vskip 1pt}
11 & 1.127\textless{}FWHM\_IMAGE\_I\_E\textless{}2.843 & 6614 & 2.03 & 3293 & 1.01 \\
\hline
\end{tabular}
\label{tab:table4}
\end{table*}

The contamination by background extragalactic sources, the inhomogeneous interstellar extinction in star-forming regions, the possible presence of colour-excesses owing to discs and accretion activity, and the low surface-gravity and extreme youth of FFPs in Orion, together make the selection of bona-fide substellar objects quite challenging. The \Euclid passbands are not specifically designed to distinguish FFPs from other types of objects. They are broader than the passbands  commonly used in ground-based surveys because they include spectral regions affected by saturated telluric water absorption. The calibrations available for this ERO study are scarce. Improved calibrations are expected in the future when \Euclid photometry and spectroscopy of benchmark ultracool dwarfs become available.  We limit the scope of this work to present a high-purity approach to select objects using \Euclid ERO catalogue and images that is anchored to the properties of the benchmarks described in the previous section. The selection cuts adopted in this work are presented in Table~\ref{tab:table4}. To arrive at these selection cuts, we calculated the 1$\,\sigma$ dispersion around the mean of the values for the benchmark sources and we added it to both extremes of the distribution. 

\begin{figure}[htbp!]
\centering
\includegraphics[width=\linewidth]{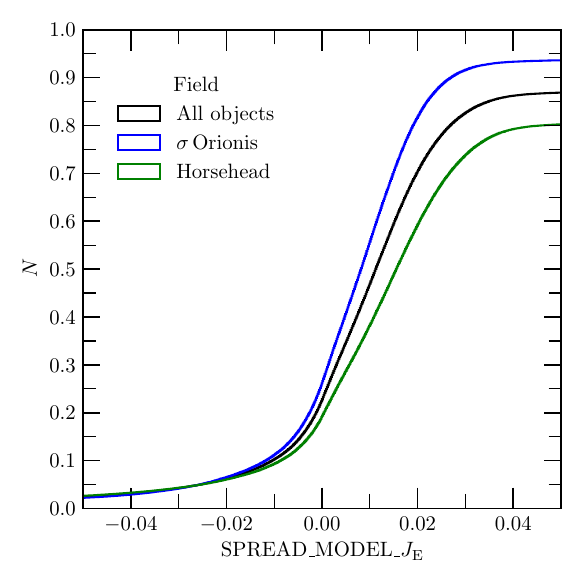}
\caption{Cumulative distribution of spread model values for the {\it Euclid\/} $\JE$ band. The distribution of values in the low-reddening part corresponding to the $\sigma$\,Orionis cluster is sharper than in the high-reddening part. }
\label{fig:spread}
\end{figure}

The ERO-SOri photometric catalogue was filtered using the cuts provided in Table~\ref{tab:table4}. The number of objects left after each step and the percentage with respect to the original sample are also given in the table. Note that the percentage of sources detected in the $\JE$ band in the whole FoV adds up to 76.83\% of the total number of sources in the ERO catalogue. Sources not detected in the $\JE$ band have not been considered in this work because young substellar objects are expected to be much brighter in the $\JE$ band than in the $\IE$ band. The CLASS$_-$STAR classifier was found to be redundant with the SPREAD$_-$MODEL parameter, and the latter was chosen because the values for the benchmarks are more stable. The FoV was divided in two regions separated by a constant RA value of \ang{85.1875;;}. We call the low-reddening/dark-background part the $\sigma$\,Orionis region (${\rm RA}<\ang{85.1875;;}$) and the high-reddening/bright-background part of the Horsehead region. We compared the distribution of SPREAD$_-$MODEL$_-$J between these two parts of the FoV and found that it is narrower in the $\sigma$\,Orionis region than in the Horsehead region (Fig.~\ref{fig:spread}). This is likely due to the influence of a brighter background on the Horsehead side owing to light reflected in the nebulosity. 
Thus, we consider that the cuts defined in this work are valid only for regions with low interstellar background and negligible extinction. For the regions affected by high sky background it will be important to obtain a new sample of substellar benchmarks using the \Euclid NISP spectra. 

\subsection{Selection of new substellar member candidates in the \texorpdfstring{$\sigma$\,Orionis}{sigma Orionis} cluster with \Euclid data}
\label{sc:selection}

\begin{figure*}[htbp!]
\centering
\includegraphics[angle=0,width=\linewidth]{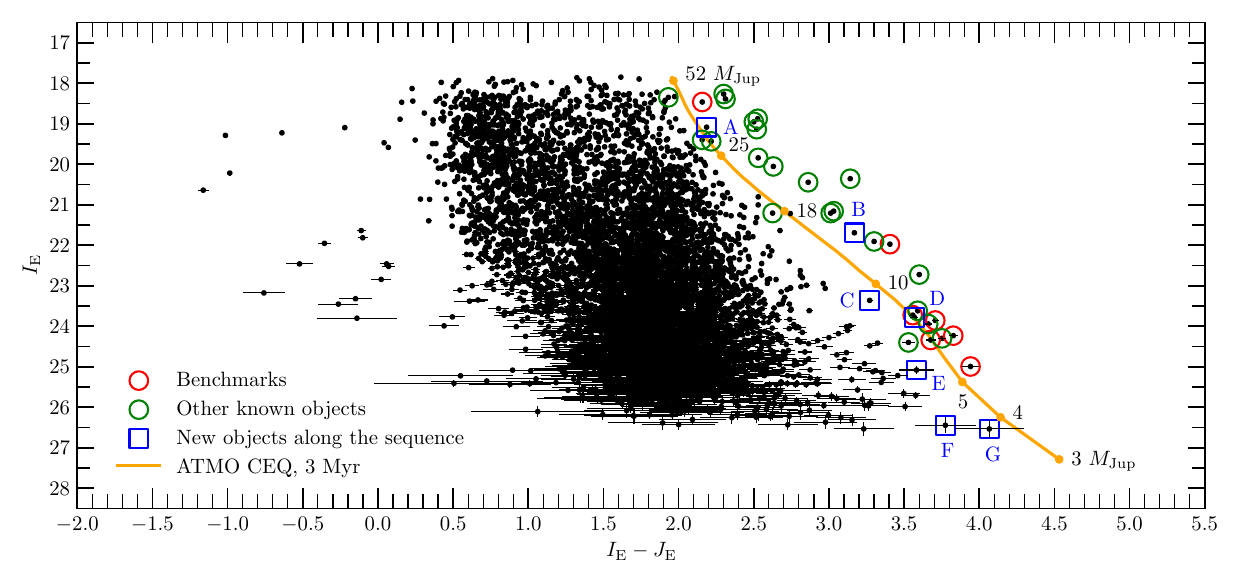}
\caption {The $\IE$ versus $\IE-\JE$ colour-magnitude diagram for the $\sigma$\,Orionis part of the FoV. Black points are all the \Euclid sources that remain after applying all the cuts. Benchmark objects are denoted with red circles. New objects near the cluster sequence are denoted with blue squares and labelled with capital letters. Known  
 sources, other than the benchmarks, in the $\sigma$\,Orionis cluster are denoted with green circles. An ATMO CEQ isochrone \citep[see][]{2020Phillips} for an age of 3\,Myr and a distance of 402.74\,pc is shown. Theoretical masses are labelled on the isochrone. }
\label{fig:soricmdij}
\end{figure*}

\begin{figure*}[htbp!]
\centering
\includegraphics[angle=0,width=\linewidth]{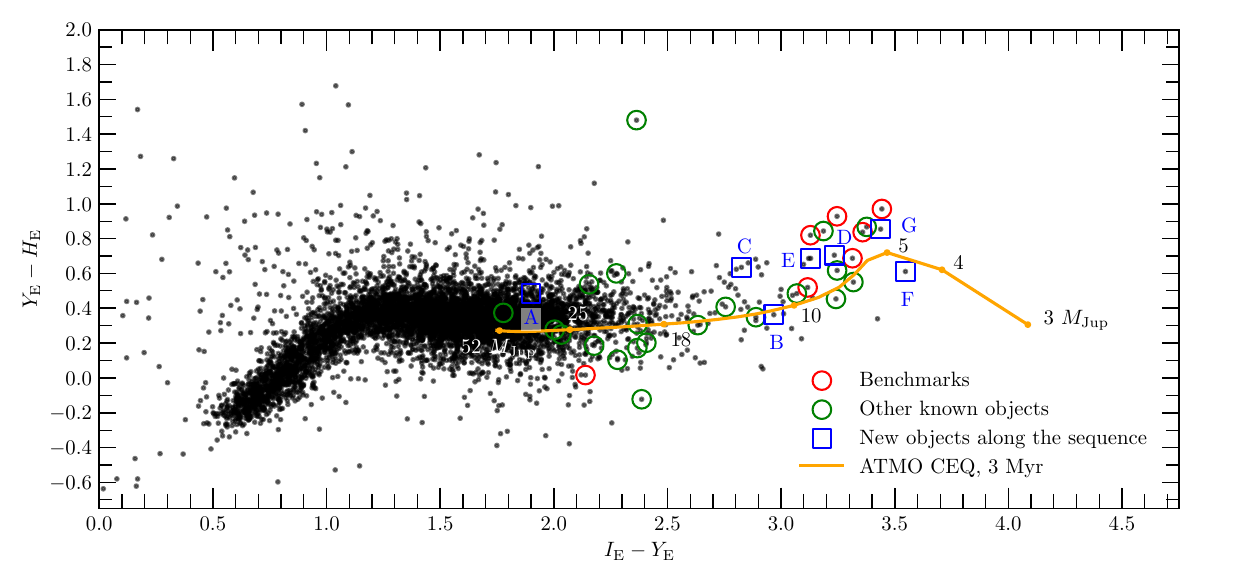}
\caption{
The $I_\mathrm{E}-Y_\mathrm{E}$ versus $Y_\mathrm{E}-H_\mathrm{E}$ %I$_E$ - Y$_E$ versus Y$_E$ - H$_E$ 
colour-colour diagram for the same region as the previous figure ($\sigma$\,Orionis). All the symbols remain the same.  
}
\label{fig:soriiyh}
\end{figure*}

\begin{figure*}[htbp!]
\centering
\includegraphics[angle=0,width=\linewidth]{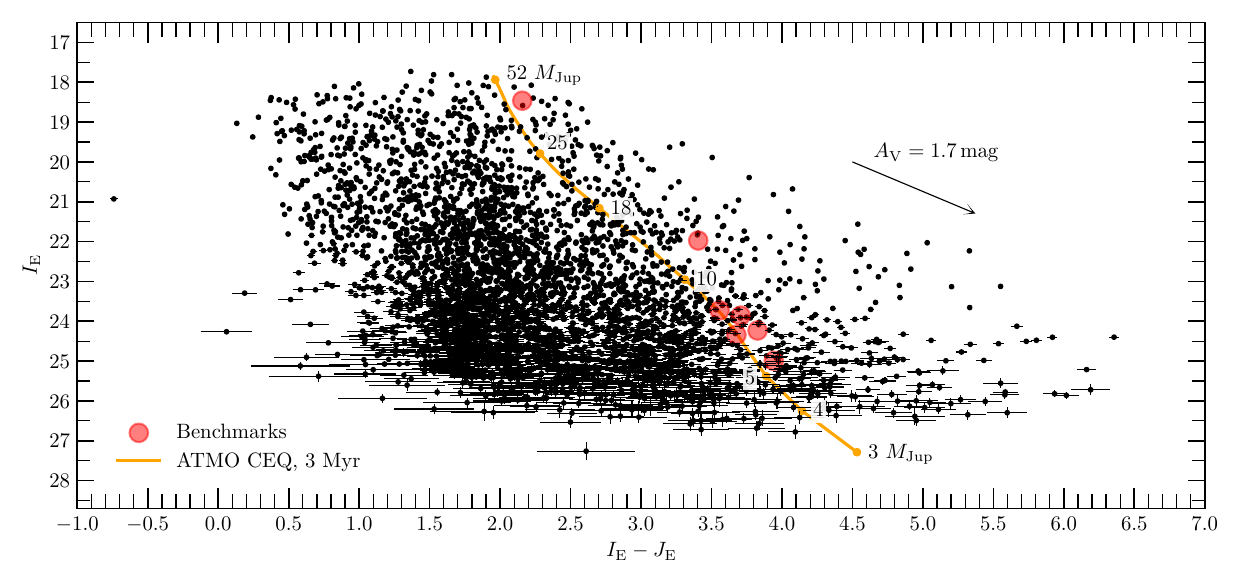}
\caption{The $I_\mathrm{E}$ versus $I_\mathrm{E}-J_\mathrm{E}$
%I$_E$ versus I$_E$ - J$_E$ 
colour-magnitude diagram for the Horsehead part of the FoV (not $\sigma$\,Orionis). The $\sigma$\,Orionis benchmarks are denoted with red circles. A reddening vector is shown as an arrow. 
}
\label{fig:horsecmdij}
\end{figure*}

\begin{figure*}[htbp!]
\centering
\includegraphics[angle=0,width=\linewidth]{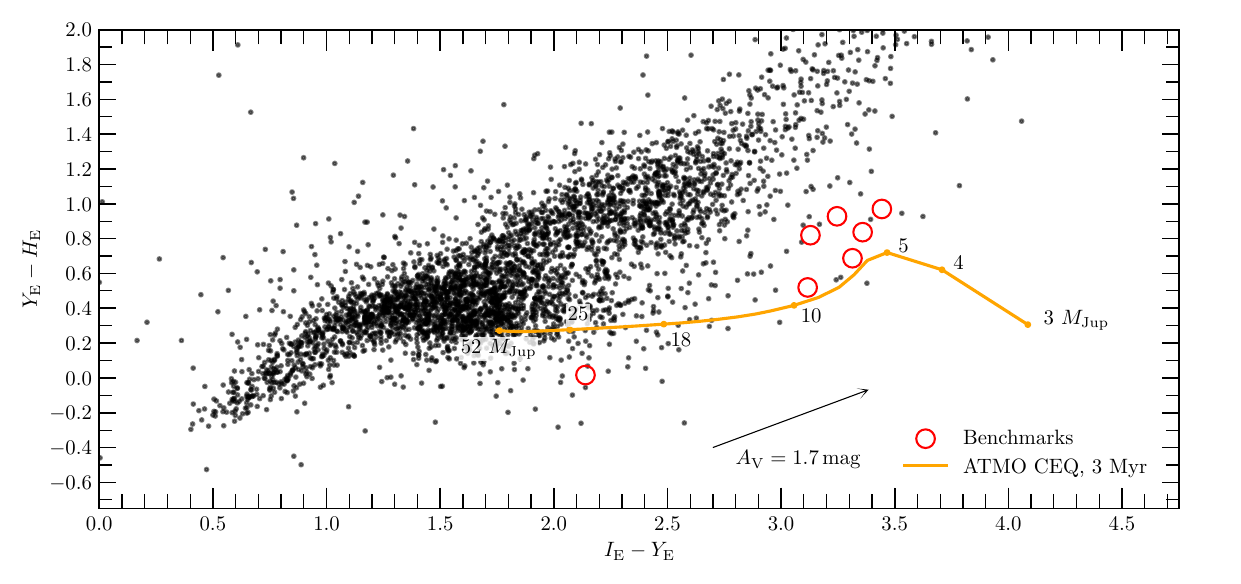}
\caption{The 
$I_\mathrm{E}-Y_\mathrm{E}$ versus $Y_\mathrm{E}-H_\mathrm{E}$
%I$_E$ - Y$_E$ versus Y$_E$ - H$_E$ 
colour-colour diagram for the Horsehead part of the FoV (not $\sigma$\,Orionis). The $\sigma$\,Orionis benchmarks are denoted with red circles. A reddening vector is shown as an arrow. 
}
\label{fig:horseiyh}
\end{figure*}

After applying all the selection cuts defined above, only 2\% of the sources in the initial catalogue remained. They are plotted in the $\IE$ versus $\IE-\JE$ colour-magnitude diagram (CMD) following the approach of previous searches for substellar objects in the $\sigma$\,Orionis region \citep{2012Ramirez}, and compared with the 3-Myr isochrone provided by the ATMO models of \cite{2020Phillips} that have been transformed into the \Euclid photometric system for this work. These models have been tested using the dynamic lithium-boundary method for brown dwarf binaries with dynamical masses and found to provide better fits to the observational data than other sets of models in the literature \citep{2022Martin}. The \Euclid data and the CEQ (equilibrium chemistry) ATMO 3-Myr isochrone are shown in the CMD displayed in Fig.~\ref{fig:soricmdij}. The benchmarks clearly define the $\sigma$\,Orionis sequence, and other objects in the \Euclid data that appear to follow this cluster sequence have been previously identified in the literature as photometric candidate members \citep{2012Ramirez}. Their \Euclid coordinates and photometry are provided in Table~\ref{tab:table5}. Seven new objects were found to be located close to the cluster sequence and well separated from the cloud of background sources, including two very faint ones that extend the sequence to fainter magnitudes than previous surveys. The three brighter objects were retrieved in the VHS catalogue and their coordinates were found to agree within $100\,{\rm mas}$, so they do not have high proper motion. 
The coordinates and photometry of these seven new \Euclid objects of interest identified in the $\sigma$\,Orionis cluster sequence are given in Table~\ref{tab:table6}. 

\begin{table*}[htbp!]
 \caption{\Euclid photometry of previously-known non-benchmark objects in the $\sigma$\,Orionis cluster sequence.}
\begin{tabular}{cc@{\hskip 2em}cc@{\hskip 2em}cc@{\hskip 2em}cc@{\hskip 2em}cc}
\hline\hline\noalign{\vskip 1pt}
RA (J2000) & \,Dec (J2000) & \IE & $\sigma$(\IE) & \YE & $\sigma$(\YE) & \JE & $\sigma$(\JE) & \HE & $\sigma$(\HE) \\ 
\hline\noalign{\vskip 1pt}
\ang{84.847645;;} & \ang{-2.682660;;} & 20.445 & 0.002  & 17.812 & 0.002  & 17.583 & 0.001  & 17.508 & 0.001  \\
\ang{84.872332;;} & \ang{-2.777320;;} & 21.207 & 0.003  & 18.451 & 0.003  & 18.196 & 0.002  & 18.043 & 0.002  \\
\ang{84.885072;;} & \ang{-2.872183;;} & 24.299 & 0.027  & 20.923 & 0.016  & 20.547 & 0.009  & 20.055 & 0.006  \\
\ang{84.943614;;} & \ang{-2.406478;;} & 21.158 & 0.003  & 18.269 & 0.003  & 18.127 & 0.002  & 17.920 & 0.002  \\
\ang{84.989081;;} & \ang{-2.835013;;} & 23.618 & 0.016  & 20.300 & 0.011  & 20.028 & 0.007  & 19.749 & 0.006  \\
\ang{85.028991;;} & \ang{-2.601451;;} & 22.725 & 0.008  & 19.478 & 0.005  & 19.124 & 0.003  & 18.861 & 0.003  \\
\ang{85.032282;;} & \ang{-2.376239;;} & 23.945 & 0.022  & 20.758 & 0.017  & 20.284 & 0.009  & 19.915 & 0.007  \\
\ang{85.076609;;} & \ang{-2.385794;;} & 20.358 & 0.002  & 17.994 & 0.003  & 17.216 & 0.001  & 16.513 & 0.001  \\
\ang{85.048215;;} & \ang{-2.859726;;} & 24.400 & 0.039  & 21.159 & 0.017  & 20.872 & 0.011  & 20.705 & 0.010  \\
\ang{84.919110;;} & \ang{-2.653550;;} & 18.348 & 0.001  & 16.570 & 0.001  & 16.416 & 0.001  & 16.196 & 0.001  \\
\ang{84.893058;;} & \ang{-2.646380;;} & 18.266 & 0.001  & 15.986 & 0.001  & 15.967 & 0.001  & 15.882 & 0.001  \\
\ang{84.815680;;} & \ang{-2.640655;;} & 18.387 & 0.001  & 16.209 & 0.002  & 16.075 & 0.001  & 16.024 & 0.001  \\
\ang{84.861922;;} & \ang{-2.615620;;} & 18.952 & 0.001  & 16.585 & 0.001  & 16.452 & 0.001  & 16.414 & 0.001  \\
\ang{85.018897;;} & \ang{-2.611691;;} & 18.877 & 0.001  & 16.491 & 0.001  & 16.351 & 0.001  & 16.613 & 0.001  \\
\ang{85.074216;;} & \ang{-2.448392;;} & 19.842 & 0.002  & 17.473 & 0.002  & 17.312 & 0.001  & 17.164 & 0.001  \\
\ang{85.142016;;} & \ang{-2.434096;;} & 20.055 & 0.002  & 17.779 & 0.002  & 17.424 & 0.001  & 17.179 & 0.001  \\
\ang{84.813607;;} & \ang{-2.364092;;} & 19.135 & 0.001  & 16.728 & 0.002  & 16.615 & 0.002  & 16.525 & 0.002  \\
\ang{84.808099;;} & \ang{-2.272735;;} & 21.907 & 0.005  & 18.837 & 0.007  & 18.606 & 0.005  & 18.353 & 0.004  \\ \hline
\end{tabular}
\label{tab:table5}
\end{table*}

\begin{table*}[htbp!]
    \caption{ \Euclid objects of interest in the $\sigma$\,Orionis region.}
\begin{tabular}{lcc@{\hskip 2em}cc@{\hskip 2em}cc@{\hskip 2em}cc@{\hskip 2em}cc}
\hline\hline\noalign{\vskip 1pt}
Code & RA (J2000)    & \,Dec (J2000)   & \IE      & $\sigma$(\IE) & \YE      & $\sigma$(\YE) & \JE      & $\sigma$(\JE) & \HE      & $\sigma$(\HE) \\ 
\hline\noalign{\vskip 1pt}
A    & \ang{85.162830;;} & \ang{-2.292707;;} & 19.084 & 0.001  & 17.184 & 0.002  & 16.898 & 0.001  & 16.699 & 0.001  \\
B    & \ang{84.807391;;} & \ang{-2.529354;;} & 21.692 & 0.007  & 18.724 & 0.006  & 18.523 & 0.004  & 18.361 & 0.004  \\
C    & \ang{85.036499;;} & \ang{-2.714769;;} & 23.362 & 0.019  & 20.536 & 0.014  & 20.091 & 0.007  & 19.901 & 0.007  \\
D    & \ang{85.119119;;} & \ang{-2.851646;;} & 23.779 & 0.025  & 20.545 & 0.011  & 20.211 & 0.007  & 19.840 & 0.005  \\
E    & \ang{84.892801;;} & \ang{-2.981698;;} & 25.081 & 0.110  & 21.951 & 0.067  & 21.499 & 0.037  & 21.264 & 0.030  \\
F    & \ang{85.146780;;} & \ang{-2.918538;;} & 26.444 & 0.194  & 22.897 & 0.095  & 22.670 & 0.061  & 22.286 & 0.042  \\
G    & \ang{84.818225;;} & \ang{-2.395095;;} & 26.536 & 0.221  & 23.098 & 0.134  & 22.469 & 0.062  & 22.243 & 0.051  \\ \hline
\end{tabular}
\label{tab:table6}
\end{table*}

Further examination of the cluster sequence, its degree of agreement with the ATMO isochrone and the location of new candidate members was made in the colour-colour diagram shown in Fig.~\ref{fig:soriiyh}. The coolest benchmarks define a well separated locus away from the cloud of contaminating sources. The behaviour of the benchmarks is qualitatively fairly well reproduced by the isochrone, although quantitatively the fit could be improved because the isochrone does not reach as large $\YE-\HE$ colour as observed. The blueing of the isochrone in the $\YE-\HE$ colour beyond $\IE-\YE \ge$3.5 is an effect of the appearance of methane in the transition from L to T-type spectra. 
The new sources with codes D, E, and G fall within the L-type benchmark locus, making them strong FFP candidates. Source F is slightly bluer in $\YE-\HE$ than the faintest benchmark, and is also closer to the isochrone, suggesting that it might be the first L/T transition FFP identified in the $\sigma$\,Orionis cluster. Confirmation of these tentative assessments requires spectroscopy.   

To check the effects of reddening in the selection of substellar candidates, the same cuts that were applied to the $\sigma$\,Orionis region were also applied to the Horsehead regions. The CMD and colour-colour diagrams are shown in Figs.~\ref{fig:horsecmdij} and \ref{fig:horseiyh}, respectively. The separation between the cluster sequence defined by the benchmarks and the cloud of sources is no longer well defined in the CMD and the locus of benchmarks in the colour-colour diagram is not well isolated. This example shows the difficulties of selecting substellar candidates in regions with high interstellar reddening. Future work will address this issue.

\begin{figure*}[htbp!]
\centering
\includegraphics[angle=0,width=\linewidth]{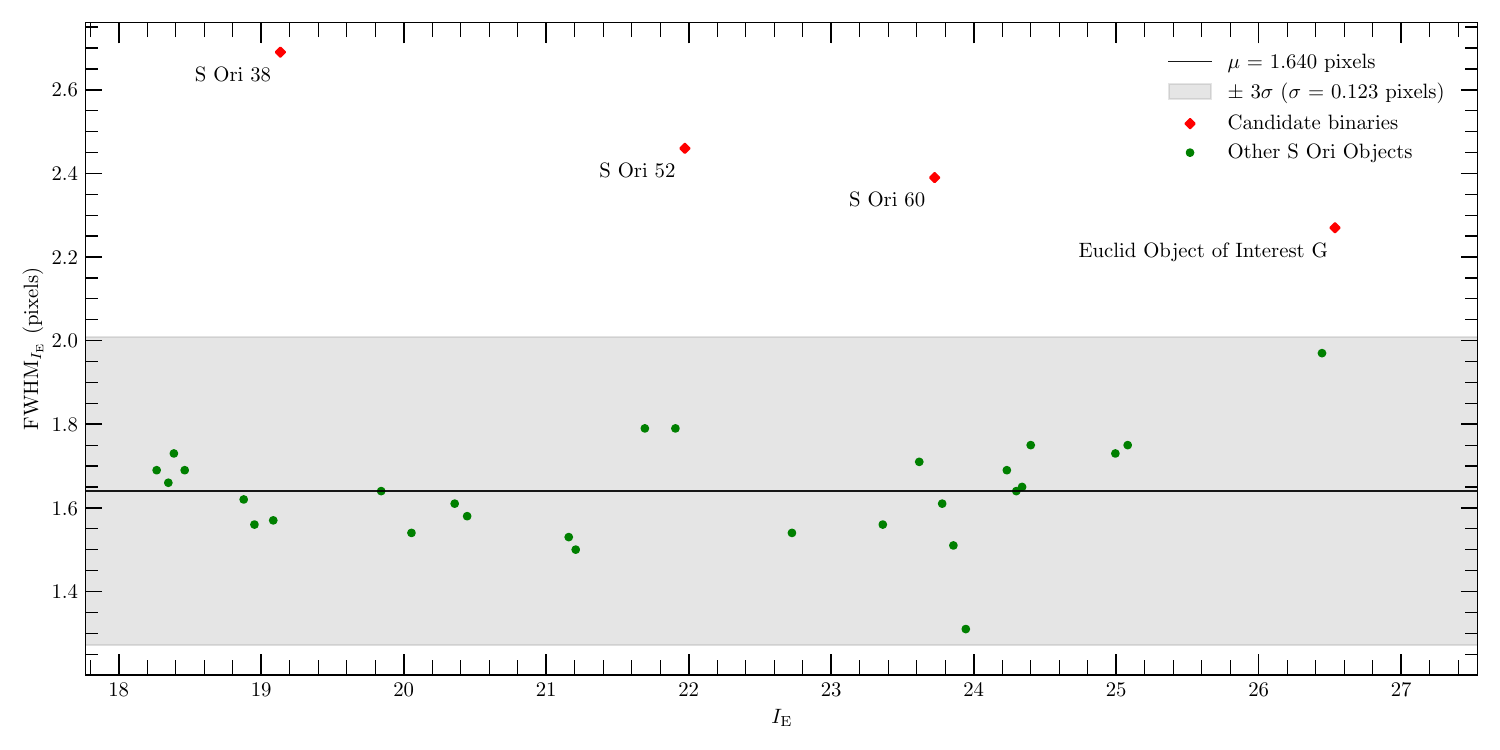}
\caption{FWHM (in pixels) versus \IE apparent magnitude (aperture photometry) for objects confirmed by \Euclid to be in the $\sigma$\,Orionis cluster sequence. Both the FWHM and \IE photometry values come from the ERO catalogue and were measured in the \Euclid VIS images. Four objects were found to have FWHM values larger than the mean value of cluster members (more than 3$\,\sigma$\, significance) and hence are considered as possible binaries that deserve further scrutiny. Two of them are benchmark objects, namely (S\,Ori\,52 and S\,Ori\,60), one is a  known object (S\,Ori\,38), and the faintest one is a new discovery (\Euclid object of interest G). 
}
\label{fig:FWHM_MAG_I}
\end{figure*}

To check for the presence of binaries, we show in Fig.~\ref{fig:FWHM_MAG_I} the FWHM values (in pixels) versus aperture magnitudes measured in the VIS images for all the objects under study (benchmarks, confirmed candidates in the cluster sequence, and new discoveries). Besides the two binary candidates among the benchmarks, there is one more candidate among the confirmed objects and one more among the new ones found with \Euclid. The object labelled as G could be the first $\sigma$\,Orionis counterpart to the Jupiter-mass binary candidates reported in the Trapezium cluster \citep{McCaughrean}, but needs confirmation with higher spatial resolution images that could be provided by HST optical imaging observations.   

\section{The \Euclid substellar IMF of the \texorpdfstring{$\sigma$\,Orionis}{sigma Orionis} cluster}\label{sc:IMF}

The results reported in this work are useful to revise the very low-mass IMF of the $\sigma$\,Orionis cluster and try to extend it deeper into the planetary-mass regime. The \textit{Gaia} sample covers the domain of very low-mass stars and \Euclid provides a continuation into the substellar-mass regime, reaching down to about $4\,M_{\rm J}$. %An $\IE$ versus $\IE - \HE$ colour-magnitude diagram of the sequence of confirmed cluster members together with the new photometric candidates and the ATMO CEQ 3-Myr isochrone is provided in Fig.~\ref{fig:soricmdih}. The same figure shows the number of objects expected per magnitude bin that we have estimated using a log-normal IMF. This shape of the IMF has been proposed to be a universal outcome of star formation \citep{1996Adams}. The IMF has been normalised to the number of objects at the peak of the IMF using a catalogue of Gaia DR3 members \citep{2023Zerjal}. The number of substellar objects reported in this work is much larger than those predicted by a log-normal IMF. Such an IMF can be ruled out with high confidence in the substellar domain.  

\begin{figure}[htbp!]
\centering
\includegraphics[angle=0,width=\linewidth]{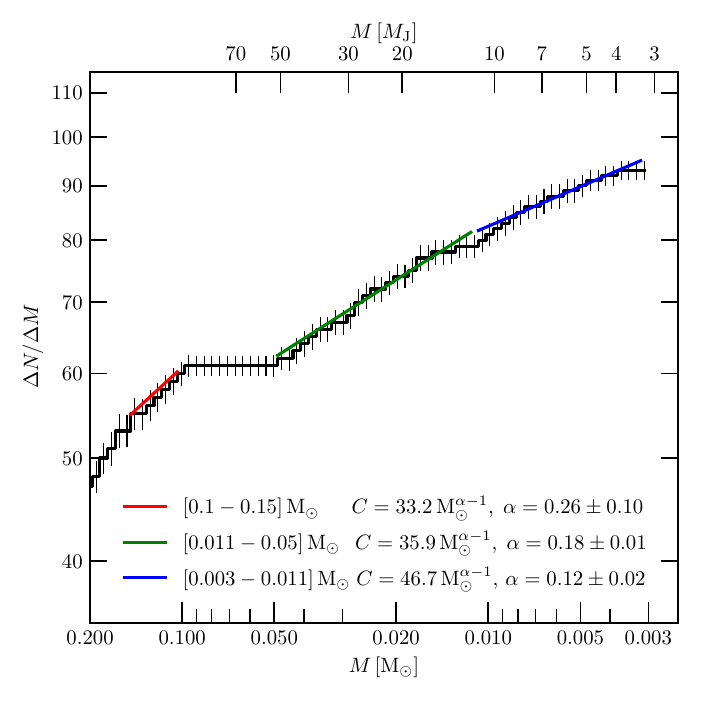}
\caption{Combined very low-mass \Euclid-\textit{Gaia} IMF of the \texorpdfstring{$\sigma$ Orionis}{sigma Orionis} cluster with linear fits in three different mass regions. 
}
\label{fig:IMF}
\end{figure}

The mass-luminosity relationship from the 3-Myr ATMO CEQ models has been used for the substellar domain. The PAdova and TRieste Stellar Evolution Code (\texttt{PARSEC}) models \citep{2012MNRAS.427..127B, 2020MNRAS.498.3283P} were used for the stellar domain. 
The IMF of the $\sigma$\,Orionis cluster using the \textit{Gaia}-DR3 membership study by \cite{2023Zerjal}, combined with the results of this work, is displayed in Fig.~\ref{fig:IMF}. Our results are consistent with a multi-power-law distribution for the IMF. 

 A comprehensive study of the IMF within a distance of 20\,pc from the Sun has reported a change in the slope in the substellar domain \citep{2024Kirk}. Those authors claimed that the Solar vicinity IMF can be expressed as $\diff N/\diff M = C\,M^{-\alpha}$ with four different values of the power-law exponent for different mass intervals. In particular, in this work we are concerned with the low-mass tail of the IMF where the slope estimated by \cite{2024Kirk} steepens from a value of $\alpha =$ 0.25 in the mass range $0.05\,M_\odot<M<0.22\,M_\odot$ to $\alpha =$ 0.60 in the mass range $0.01\,M_\odot< M <0.05\,M_\odot$. 
 
 In this study, we identify three different mass regimes: the very low-mass stellar domain from 0.15 to 0.1 M$_\odot$ with $\alpha =$ 0.26$\pm$0.10; the brown dwarf domain from 0.1 to 0.011 M$_\odot$ with $\alpha =$ 0.18$\pm$0.01; and the planetary-mass domain from 0.011 to 0.003 M$_\odot$ with $\alpha =$ 0.12$\pm$0.02. These values have been obtained with linear fits that are shown in Fig.~\ref{fig:IMF}. We excluded from the fits the mass range between 0.1 M$_\odot$ and 0.05 M$_\odot$ because those objects are too faint to be complete for Gaia and too bright for \Euclid . Error bars quoted for the IMF slopes were estimated by simulations of the effects coming from age, distance and photometric uncertainties.  
 
Our $\sigma$\,Orionis IMF results are consistent with the field in the very low-mass stellar regime and extend deeper into the substellar regime than the field IMF. We do not  confirm a steepening of the substellar IMF at the planetary-mass end. These comparisons are affected by low number statistics. The census of directly imaged FFPs should be increased significantly to investigate the possibility of substellar IMF variations in different environments that could be an indication of specific formation pathways in the planetary-mass domain. These results demonstrate that \Euclid can play a significant role in the detailed study of the low-mass shape of the IMF and particularly in shedding light on the formation mechanisms of FFPs. Detailed theoretical models developed by different groups have indicated that the shape of the IMF is a useful indicator of the dominant mode of star formation in a given region \citep{1996Adams,2005Chabrier,2015Thies}, and that a multi-power-law IMF could arise from the interplay between the mass-dependence and the time-dependence of exponential growth in a distribution of accreting protostars \citep{2020Essex}.  

\section{Final remarks: The impact of \Euclid on the study of FFPs in star-forming regions\label{sc:final} }

This work is a showcase of the power of the  
 \Euclid mission to provide the area and depth required to explore the very low-mass population, including FFPs of nearby star-forming regions and very young open clusters. In particular, for the well-known $\sigma$\,Orionis cluster, we show that the sensitivity of the \Euclid images is capable of probing down to FFPs that could have masses as low as $4\,M_{\rm J}$ according to theoretical models (for 3\,Myr ages) and at a distance of 400\,pc. This potential could be compromised by severe contamination from numerous background extragalactic sources if we do not use stringent selection procedures. Using the \Euclid data for seven benchmark objects in $\sigma$\,Orionis, we have developed a high-purity method to filter out the contamination. This method is valid for regions of low reddening, but it needs additional work to be generalised to regions with any reddening. We note that the \Euclid NISP spectra will likely play an important role in this effort. 
 Additionally multi-epoch observations with \Euclid during the lifetime of the mission, possibly filling gaps in the cosmological surveys, can enable the study of proper motions and photometric variability that are useful probes for the study of the low-mass population in star-forming regions.  
 
 This is the first of a series of papers that intend to explore several star-forming regions using \Euclid observations. The observations presented here provide a glimpse of the power of \Euclid to shed light on the long-standing question of the putative low-mass cutoff of the IMF predicted long time ago by the theory of opacity-limited fragmentation and collapse of molecular clouds. 
 Our IMF for the $\sigma$\,Orionis open cluster extends previous studies to lower planetary masses and suggests that there could be a difference in slope in the substellar regime between this young open cluster and the field, hinting at a possible sensitivity to environmental conditions. This study demonstrates the great potential of \Euclid to tackle the study of the substellar IMF in nearby star-forming regions and very young open clusters.

\begin{acknowledgements}
 % We thank XXX and YYY for helpful comments on the
 % manuscript. 
 This work has made use of the Early Release Observations (ERO) data from the \Euclid mission of the European Space Agency (ESA), 2024. We thank I. Baraffe and M. Phillips for making available a digitised version of their isochrones in the \Euclid passbands. 
  E.L.M. and M.{\v Z}. are supported by the European Research Council
  Advanced grant SUBSTELLAR, project number 101054354. This research has made use of the Spanish Virtual Observatory (https://svo.cab.inta-csic.es) project funded by MCIN/AEI/10.13039/501100011033/ through grant PID2020-112949GB-I00 at Centro de Astrobiolog\'{i}a (CSIC-INTA). DB and NH have been supported by PID2019-107061GB-C61 by the same agency. P.M.B. is funded by Instituto Nacional de T\'ecnica Aeroespacial through grant PRE-OVE. P.C. acknowledges financial support from the Spanish Virtual Observatory project (grant PID2020-112949GB-I00). N.P.B. is funded by Vietnam National Foundation for Science and Technology Development (NAFOSTED) under grant number 103.99-2020.63.
  NL and VJSB acknowledge financial support from the Agencia Estatal de Investigaci\'on (AEI/10.13039/501100011033) of the Ministerio de Ciencia e Innovaci\'on and the ERDF `A way of making Europe' through project PID2022-137241NB-C1. CdB acknowledges support from a Beatriz Galindo senior fellowship (BG22/00166) from the Spanish Ministry of Science, Innovation and Universities. 
\AckEC
\end{acknowledgements}

\bibliography{Euclid,EROplus}

%\begin{appendix}
 % \onecolumn %If you don't want single column for the Appendix,  
%\section{First part of the Appendix\label{apdx:A}}
%Note that appendices in A\&A come {\it after\/} the references.

%\newpage

%\begin{figure*}[htbp!]
%\centering
%\includegraphics[angle=0,width=\linewidth]{CMD_I_H_vs_I_SOri_AUTO_1STD.PDF}
%\caption{The 
%$I_\mathrm{E}$ versus $I_\mathrm{E}-H_\mathrm{E}$
%I$_E$ versus I$_E$ - H$_E$ 
%colour magnitude diagram for the $\sigma$\,Orionis region. The expected number of $\sigma$\,Orionis members per magnitude bin estimated by us using a standard IMF are provided. See text for discussion.  
%}
%\label{fig:soricmdih}
%\end{figure*}

%\begin{figure}[htbp!]
%\centering
%\includegraphics[angle=0,width=1.0\hsize]{HJ-zoom-bin15.png}
%\caption{A wide binary planetary-mass candidate found with  {\it Euclid\/} NISP in the Taurus ERO field. H-band to the left and J-band to the right. Note that there are spurious bright pixels and residuals from the spectroscopic exposure.}
%\label{fig:bin}
%\end{figure}

%\section{Second part of the Appendix, i.e., Appendix B}
%
 
%\clearpage

%\end{appendix}
\label{LastPage}
\end{document}